\documentclass[12pt]{iopart}

\usepackage{iopams}
\usepackage{setstack,amssymb,graphicx,color,subfigure}

\renewcommand{\mbox}{\rm}

\newcommand{\bp}{\boldsymbol{p}}
\newcommand{\bq}{\boldsymbol{q}}
\newcommand{\bJ}{\boldsymbol{J}}
\newcommand{\bt}{\boldsymbol{\theta}}
\newcommand{\bm}{\boldsymbol{m}}
\newcommand{\bzero}{\boldsymbol{0}}
\newcommand{\nume}{\nu}
\newcommand{\denom}{\mu}

\begin{document} 

\title{On algebraic damping close to inhomogeneous Vlasov equilibria
  in multi-dimensional spaces}

\author{Julien Barr\'e$^{1}$ and Yoshiyuki Y Yamaguchi$^{2}$}
\address{$^1$ Laboratoire J.A. Dieudonn\'e, Universit\'e de Nice Sophia-Antipolis, UMR CNRS 7351, Parc Valrose, F-06108 Nice Cedex 02, France}
\address{$^2$ Department of Applied Mathematics and Physics, Graduate School of Informatics, Kyoto University, Kyoto 606-8501, Japan}
\ead{jbarre@unice.fr}

\begin{abstract}
    We investigate the asymptotic damping of a perturbation 
    around inhomogeneous stable stationary states
    of the Vlasov equation in spatially multi-dimensional systems.
    We show that branch singularities of the Fourier-Laplace transform
    of the perturbation yield algebraic dampings.
    In two spatial dimensions, we classify the singularities and compute the associated
    damping rate and frequency. This 2D setting also applies to spherically symmetric self-gravitating systems.
    We validate the theory using a toy model and an advection equation associated with the isochrone model, 
    a model of spherical self-gravitating systems.

\end{abstract}
\submitto{\JPA}
\maketitle

\section{Introduction}

The Vlasov equation, often called Collisionless Boltzmann equation,
describes, over a certain time frame, the large
scale dynamics of Hamiltonian systems of interacting particles, in the
limit where each particle feels the effect of many others. It is thus
found in many fields, among which plasma physics of course, and
astrophysics, where it describes self-gravitating systems.

The Vlasov equation shows a notoriously rich 
dynamics. First, it possesses a continuous infinity of stationary states,
whose study may in itself be a complicated problem. The next problem,
the study of the linearized dynamics close to a stationary state, has
a long story. In 1946, Landau~\cite{Landau46} formally showed that
close to a stable homogeneous stationary state, a density perturbation
may decay exponentially, which was a very surprising result for a
Hamiltonian system. This was the starting point of an extremely
abundant physical literature. On the mathematical side, it has been
proved that Landau analysis is correct \cite{Maslov85,Degond86}. It is
also known that the exponential damping fails when the reference state
or the perturbation is not analytic~\cite{Weitzner67}, or the spatial
domain unbounded~\cite{Glassey95}.

The non-linear case, that is the study of a perturbation close to a
homogeneous stationary state, under the full Vlasov dynamics, is much
more complicated. The subject has witnessed spectacular progresses
recently~\cite{MouhotVillani,MouhotVillani2,Lin10}.

Most studies on Landau damping deal with homogeneous stationary
states.  By comparison, the literature on perturbations of 
inhomogeneous states is less developed, although these states are very
important.  Despite the technical difficulties involved, a lot of work has
been done in the astrophysical literature.  Kalnajs \cite{Kalnajs77},
and then Polyachenko and Schuckman \cite{Polyachenko81} have developed
a powerful formalism, sometimes called the "matrix method", to solve
the linearized Vlasov equation in a inhomogeneous context. Since
then, it has been used to study the instability of many models of
stellar systems, and compute the growth rates
(see~\cite{BinneyTremaine} for a textbook account).  Some other
methods to compute instability rates have been introduced and applied
to toy models recently~\cite{Jain07,Campa10,Bachelard10}. Purely
oscillating modes in 1D have been investigated
in~\cite{Mathur90}. Also, when the question is to prove stability and
not to compute a decay rate, there exist powerful variational
methods~\cite{BinneyTremaine}.  However, computing decay rates for
non-oscillating stable situations is more tricky than computing
instability rates, since it involves an additional analytic continuation,
as in Landau's original analysis.  In the astrophysical literature, we
are aware of only two papers performing this continuation numerically,
and thus explicitly computing the analog of Landau damping
rates~\cite{Weinberg94,Weinberg00}. In \cite{BOY10}, the continuation
is performed analytically, but on a one dimensional toy model.  Thus,
although Landau damping is considered an important player in the
dynamics of stellar systems (see for instance \cite{BinneyTremaine}),
there are very few actual computations of damping rates close to
inhomogeneous stationary states.

In addition, these studies done in the astrophysical context do not
mention a fundamental difference between homogeneous and
inhomogeneous stationary states: in the inhomogeneous case, no
matter what is the regularity of the stationary state and the
perturbation, the asymptotic linear decay is never exponential. This
has been seen in the Kuramoto model, which shares some properties with
the Vlasov equation~\cite{Strogatz}, and on a type of Vlasov equation
in \cite{Smereka}. The picture is as follows: the dynamics may show a
transient exponential decay governed by a Landau pole, but the
asymptotic decay is always algebraic. This phenomenology is also well
known for the 2D Euler equation~\cite{Case1960}, which is in many
respects similar to the Vlasov equation.
We note that such algebraic decays may also arise close to a homogeneous
state, when the perturbation or the reference state is not regular
enough~\cite{Weitzner67}; we stress that the situation is
different close to inhomogeneous states, because the algebraic decay
occurs for all reference states and perturbations. \cite{BOY11}
contains a detailed analysis of this phenomenon in 1D, including
derivation of the decay exponents and comparison with direct numerical
simulations. In this paper, we extend the analysis of~\cite{BOY11} to
multidimensional systems, putting our emphasis on 2D systems and 3D
systems with spherical symmetry. These includes some important models
of stellar systems.

More specifically, we first use the standard matrix method to formally
solve the linear dynamics in a Laplace transformed
space (section~\ref{sec:matrixmethod}). From this starting point:\\
i) we identify and classify the singularities appearing in the Laplace
transform of the perturbation. It turns out that the zoology of
singularities in two dimensions is much richer than in one dimension,
see section~\ref{sec:singularities-FG}.\\
ii) we exhibit the asymptotic decay for each type of singularities
in section~\ref{sec:asymptotic}.\\
iii) we introduce advection equations associated with the linearized
Vlasov equations in section~\ref{sec:preparation-numerics}.
These advection equations have a solution in integral form;
hence the numerical task to obtain the temporal evolution
of the system is reduced from solving the (linearized) Vlasov equation
in $2d$-dimensional phase space to performing integrals
in a $d$-dimensional space, where $d$ is the spatial dimension.\\
iv) Using the above theory, we analyze the asymptotic decay
of a perturbation in a toy model and a spherically symmetric stellar model,
and check the theory against the numerically computed exact solutions
to the simple advection models,
see sections~\ref{sec:toy_model},~\ref{sec:self-grav}.\\

Our analysis of the asymptotic decay of a perturbation is formal, and
moreover relies on the linearized Vlasov equation. There is no
guarantee that this is relevant to understand the asymptotic decay of
the non-linear Vlasov equation. In principle, our results should then
be supplemented by direct numerical simulations of the full Vlasov
equation. However, such simulations are difficult in more than
one spatial dimension, since we are aiming at an asymptotic in time
regime, while keeping a good spatial precision. We left this study for
future work. This is why we analyze instead with the present theory
easily solvable linear advection equations, for illustrative purpose.

\section{Solution to the linearized Vlasov equation: the matrix method}
\label{sec:matrixmethod}
Throughout the paper, we will use bold letters
($\bq,~\bp,~\bt,~\bJ,~\ldots$) to denote vectors.

\subsection{System}
\label{sec:system}
We start with the Vlasov equation in $d$ spatial dimensions for the one-particle
distribution function $f(\bq,\bp,t)$,
\begin{equation}
    \label{eq:vlasov}
    \partial_{t}f + \nabla_{\bp}H \cdot \nabla_{\bq}f
    - \nabla_{\bq} H \cdot \nabla_{\bp}f = 0, 
\end{equation}
where $\bq \in X\subset \mathbb{R}^d$ is the position variable, 
$\bp\in\mathbb{R}^d$ the conjugate momentum variable,
and $\nabla_{\bq},~\nabla_{\bp}$ denote respectively
the gradients with respect to $\bq$ and $\bp$.
The domain $X$ is $\mathbb{R}^{3}$ in a stellar system,
and $[0,2\pi)^{d}$ if the system has periodic boundary conditions.
$H$ is the one-particle Hamiltonian defined by
\begin{equation}
    \label{eq:one-particle-hamiltonian}
    H[f](\bq,\bp,t) = \frac{\bp^{2}}{2} + \Phi[f](\bq,t)
    + \Phi_{\mbox{ext}}(\bq).
\end{equation}
The potential $\Phi[f](\bq,t)$ is defined from 
the two-body interaction potential $v$ and the distribution $f$ as
\begin{equation}
    \label{eq:potential}
    \Phi[f](\bq,t) = \int_{\mathbb{R}^d} d\bp
    \int_{X} d\bq~ v(\bq-\bq') f(\bq',\bp,t).
\end{equation}
The external potential $\Phi_{\mbox{ext}}(\bq)$ creates a force
$F_{\mbox{ext}}(\bq)$:
\begin{equation}
    \label{eq:external-force}
    F_{\mbox{ext}}(\bq)=-\nabla_{\bq}\Phi_{\mbox{ext}}(\bq).
\end{equation}

Let $f_{0}$ be a stationary solution to the Vlasov equation~(\ref{eq:vlasov}).
For the stationary solution, 
the potential $\Phi[f_{0}]$ and the one-particle Hamiltonian $H[f_{0}]$ 
are independent of time.
In this paper, we limit ourselves to situations where the one-particle
Hamiltonian $H[f_{0}]$ is integrable.
Thus, we may introduce actions $\bJ=(J_1,\ldots,J_d)$
and conjugate angles $\bt=(\theta_1,\ldots,\theta_d)$,
and the one-particle Hamiltonian is a function of
the actions only, i.e. $H[f_{0}](\bJ)$.
A practically important example is given by spherically symmetric
stellar systems, see section~\ref{sub:example}.
A stationary solution can be constructed by taking $f_{0}$
as a function of actions, satisfying self-consistent conditions,
since the actions depend on $f_{0}$ through the potential $\Phi[f_{0}]$.

We now consider a small
perturbation to a stationary solution $f_{0}(\bJ)$:
\begin{equation}
    f(\bt,\bJ,t) = f_{0}(\bJ)+f_{1}(\bt,\bJ,t)
\end{equation}
and write the linearized Vlasov equation for $f_{1}$: 
\begin{equation}
    \partial_{t} f_{1} +\bOmega(\bJ)\cdot\nabla_{\bt} f_{1} - 
    \nabla_{\bJ}f_{0} \cdot \nabla_{\bt} \Phi_{1} =0~,
    \label{eq:eqlin}
\end{equation}
where $\Phi_{1}=\Phi[f_{1}]$ is the perturbed potential
\begin{equation}
    \Phi_{1}(\bq,t)
    =\int_{\mathbb{R}^d} \int_{X} v(\bq-\bq')
    f_{1}(\bq',\bp,t) d\bq' d\bp,
    \label{eq:phi1}
\end{equation}
and $\bOmega(\bJ)=\nabla_{\bJ}H[f_{0}](\bJ)$ is the vector of
the frequencies in the unperturbed potential.

\subsection{Fourier-Laplace transform}
\label{sec:fourier-laplace-transform}
To analyze the linearized Vlasov equation (\ref{eq:eqlin}),
we introduce the Fourier-Laplace transform $\hat{u}(\bm,\bJ,\omega)$ 
of a function $u(\bt,\bJ,t)$ as
\begin{equation}
    \hat{u}(\bm,\bJ,\omega)
  = \int_{-\pi}^{\pi}d\bt~ e^{-i\bm\cdot\bt}
  \int_{0}^{+\infty}dt~ e^{i\omega t}u(\bt,\bJ,t)
\end{equation}
where $\bm=(m_{1},\dots,m_{d})$ is a $d$-uplet of integers,
$\bm\cdot\bt$ is the Euclidian inner product
and ${\mbox Im}(\omega)$ is large enough to ensure convergence
of the integral with respect to $t$.
The inverse transform is then
\begin{equation}
  \label{eq:FourierLaplace_inverse}
  u(\bt,\bJ,t)=\frac{1}{(2\pi)^{d+1}}\sum_{\bm\in\mathbb{Z}^{d}}
  e^{i\bm\cdot\bt}
  \int_{\Gamma}d\omega~e^{-i\omega t}~\hat{u}(\bm,\bJ,\omega) 
\end{equation}
where $\Gamma$ is a Bromwich contour running from
$-\infty+i\sigma$ to $+\infty+i\sigma$,
and the real value $\sigma$ is larger
than the imaginary part of any singularity of 
$\hat{u}(\bm,\bJ,\omega)$ in the complex $\omega$-plane.

Simple and standard algebraic manipulations on (\ref{eq:eqlin})
yield the following expression for the Fourier-Laplace transform of
the perturbation:
\begin{equation}
    \label{eq:f1-fourier-laplace}
    \hat{f}_{1}(\bm,\bJ,\omega) = A(\bm,\bJ,\omega) 
    \hat{\Phi}_{1}(\bm,\bJ,\omega) + B(\bm,\bJ,\omega),
\end{equation}
where
\begin{equation}
    \label{eq:A}
    A(\bm,\bJ,\omega) = \frac{\bm\cdot \nabla_{\bJ}f_0(\bJ)}
    {\bm\cdot \bOmega(\bJ)-\omega},
\end{equation}
\begin{equation}
    \label{eq:B}
    B(\bm,\bJ,\omega) = \frac{g(\bm,\bJ)}{\bm\cdot \bOmega(\bJ)-\omega},
\end{equation}
and $ig(\bm,\bJ)$ is the Fourier transform of the initial perturbation
$f_1(\bt,\bJ,t=0)$ with respect to $\bt$. Without loss of generality,
we assume
\begin{equation}
    \label{eq:g0}
    g(\bzero,\bJ)=0.
\end{equation}
Notice that we can always include a non-zero $g(\bzero,\bJ)$
in the stationary state $f_0$.

\subsection{Biorthogonal functions}
\label{sec:biorthogonal-functions}
We should now solve the two coupled equations (\ref{eq:phi1}) and
(\ref{eq:f1-fourier-laplace}) to find $f_{1}$ and $\Phi_{1}$.  The
usual strategy at this point is to introduce two families of
biorthogonal functions~\cite{Kalnajs77,Polyachenko81,CluttonBrock72}.
The computations below are standard (for a textbook reference, see for 
instance~\cite{BinneyTremaine}); we reproduce them here 
to ensure that this article is self-contained.

We expand any density function $\rho(\bq)$
on the basis $\{d_{i}(\bq)\}_{i\in I}$:
\begin{equation}
    \rho(\bq)=\sum_{i\in I}a_{i}d_{i}(\bq)
\end{equation}
and any potential function $\Phi(\bq)$
on the basis $\{u_{k}(\bq)\}_{k\in K}$:
\begin{equation}
    \Phi(\bq)=\sum_{k\in K}b_{k}u_{k}(\bq).
\end{equation}
The two index sets $I$ and $K$ satisfy $K\subset I\subset\mathbb{Z}$,
and the two basis $\{d_{i}\}_{i\in I}$ and $\{u_k\}_{k\in K}$
are chosen such that
\begin{itemize}
      \item[(i)] the two families are orthogonal to each other:
    \begin{equation}
        \label{eq:biorthogonal}
        (d_{i},u_{k}) = \int_{\mathbb{R}^d} d_{i}(\bq) \bar{u}_{k}(\bq) d\bq
        = \lambda_{k}\delta_{ik}, \quad
        (i\in I,~ k\in K)
    \end{equation}
    with $\lambda_{k}\neq 0$,
    and where $\delta_{ik}$ is the Kronecker $\delta$.
      \item[(ii)] $u_{k}$ is the potential created
    by the density distribution $d_{i}$:
    \begin{equation} 
        \label{eq:ud}
        \int_{\mathbb{R}^d}v(\bq-\bq')d_{i}(\bq')d\bq'
        = \left\{
          \begin{array}{ll}
              u_{i}(\bq) & (i\in K)\\
              0 & (i\notin K)
          \end{array}
        \right.
    \end{equation}
\end{itemize}

Assuming we have at hand two such families of biorthogonal functions,
we expand the perturbation density
\begin{equation}
    \label{eq:rho1}
    \rho_{1}(\bq,t) = \int_{\mathbb{R}^{d}} f_{1}(\bq,\bp,t) d\bp
\end{equation}
and potential as
\begin{eqnarray}
    \rho_{1}(\bq,t) &=& \sum_{i\in I} a_{i}(t) d_{i}(\bq)
    \label{eq:rho1_exp}\\
    \Phi_{1}(\bq,t) &=& \sum_{k\in K} a_{k}(t) u_{k}(\bq).
    \label{eq:phi1_exp}
\end{eqnarray}
Properties (\ref{eq:biorthogonal}) and (\ref{eq:ud}) ensure that the 
coefficients are identical for $\rho_{1}$ and $\Phi_{1}$.

Using (\ref{eq:phi1_exp}), we first compute $\hat{\Phi}_{1}$,
the Fourier-Laplace transform of $\Phi_{1}$:
\begin{equation}
    \label{eq:phi1_exp2}
    \hat{\Phi}_{1}(\bm,\bJ,\omega)
    = \sum_{k\in K} \tilde{a}_{k}(\omega) c_{k}(\bm,\bJ)
\end{equation}
where $\tilde{a}_{k}$ is the Laplace transform of $a_{k}$,
and $c_{k}(\bm,\bJ)$ is the Fourier transform of $u_{k}$:
\begin{equation}
    \label{eq:c_k}
    c_{k}(\bm,\bJ) = \int u_{k}(\bq) e^{-i\bm\cdot\bt}d\bt.
\end{equation}

Substituting (\ref{eq:phi1_exp2}) into (\ref{eq:f1-fourier-laplace}),
and performing the inverse Fourier transform,
we obtain an expression for $\tilde{f}_{1}(\bt,\bJ,\omega)$,
the Laplace transform of the perturbation:
\begin{equation}
    \label{eq:f1-laplace}
    \tilde{f}_{1}
    = \frac{1}{(2\pi)^{d}}\sum_{\bm\in\mathbb{Z}^{d}}
    \left(
      \frac{\bm\cdot \nabla_{\bJ}f_{0}}{\bm\cdot \bOmega-\omega}
      \sum_{k\in K} \tilde{a}_{k}(\omega) c_{k}(\bm,\bJ)
      + \frac{g(\bm,\bJ)}{\bm\cdot \bOmega-\omega}
    \right) e^{i\bm\cdot{\bf \theta}} .
\end{equation}
Now, we multiply (\ref{eq:f1-laplace}) by $\bar{u}_{l}(\bq)~(l\in K)$
and integrate over $d\bt d\bJ$. We compute the left hand side 
using the change of variables $(\bt,\bJ)\to (\bq,\bp)$ which 
has Jacobian~$1$, and the property (\ref{eq:biorthogonal}): the result
is $\lambda_{l}\tilde{a}_l(\omega)$. In the right hand side, performing 
the integration over $\bt$ introduces the functions 
$\bar{c}_{l}(\bm,\bJ)$. The result is
\begin{equation}  
    \lambda_{l}\tilde{a}_{l}(\omega)
    = \sum_{k\in K} F_{lk}(\omega)\tilde{a}_{k}(\omega)
    + G_{l}(\omega)
\end{equation}
with
\begin{equation}
    \label{eq:F}
    F_{lk}(\omega) = \frac{1}{(2\pi)^d}
    \sum_{\bm\in\mathbb{Z}^{d}} \int
    \frac{\bm\cdot \nabla_{\bJ}f_0(\bJ)}{\bm\cdot\bOmega(J)-\omega}
    \bar{c}_{l}(\bm,\bJ) c_{k}(\bm,\bJ) d\bJ
    \quad (l,k\in K)
\end{equation}
and
\begin{equation}
    \label{eq:G}
    G_{l}(\omega) = \frac{1}{(2\pi)^d}
    \sum_{\bm\in\mathbb{Z}^{d}} \int
    \frac{g(\bm,\bJ)}{\bm\cdot\bOmega(J)-\omega} 
    \bar{c}_{l}(\bm,\bJ) d\bJ
    \quad (l\in K).
\end{equation}
The contributions from $\bm=\bzero$ clearly vanish for $F_{lk}$,
and also for $G_{l}$, thanks to assumption (\ref{eq:g0}).
Defining the $(\sharp K)\times (\sharp K)$ matrices
$\Lambda$ and $F(\omega)=(F_{lk}(\omega))$,
where $\Lambda$ is diagonal with elements $\{\lambda_{l}\}_{l\in K}$,
and the $(\sharp K)$-dimensional vectors
$G(\omega)=(G_{l}(\omega))$ and 
$\tilde{a}(\omega)=(\tilde{a}_{k}(\omega))$, the above equations
may be rewritten in compact form:
\begin{equation}
    \label{eq:a-determine}
    [ \Lambda - F(\omega) ] \tilde{a}(\omega) = G(\omega).
\end{equation}
The formal solution for $\tilde{a}$ is
\begin{equation}
    \label{eq:a-formalsol}
    \tilde{a}(\omega) = [\Lambda - F(\omega)]^{-1} G(\omega).
\end{equation}
The equation $\det (\Lambda-F(\omega))=0$ is sometimes called the dispersion 
relation. 

\subsection{Analytic continuation}
\label{sec:analytic-continuation}

Assuming that the decay for large $\bJ$ is fast enough
in the integrands, and that
the $c_{k}(\bm,\bJ)$ are regular, expressions (\ref{eq:F}) and
(\ref{eq:G}) show that functions $F_{lk}(\omega)$ and $G_{k}(\omega)$
are analytic in the upper half plane ${\mbox Im}(\omega)>0$, since the
corresponding integrals over $\bJ$ have no singularity. The
integrands are singular however for any real $\omega$
at which $\bm\cdot\bOmega(\bJ)-\omega$ vanishes.
Thus, expressions~(\ref{eq:F})
and~(\ref{eq:G}) should be analytically continued to define the $F$'s,
$G$'s and $\tilde{a}$'s in the lower half plane ${\mbox Im}(\omega)\leq 0$.
This analytical continuation is a generalization of the 
usual Landau prescription in 1D. It is sketched in some details in the 2D 
case in section~\ref{sec:abstract_pbm}.

The last step to compute the evolution of the perturbation is to 
perform an inverse Laplace transform on the functions $\tilde{a}(\omega)$.
The large time behaviour of $a_{k}(t)$ is determined by the singularities
of $\tilde{a}_{k}(\omega)$: our goal now is then to study and classify the 
singularities of $\tilde{a}(\omega)$.

\subsection{Singularities of $\tilde{a}(\omega)$ and roots of the 
dispersion relation}
\label{sec:singularities}
The singularities of $\tilde{a}$ may come
(i) from the roots of the dispersion relation $\det(\Lambda-F(\omega))=0$,
and (ii) from singularities of the functions $F$'s and $G$'s themselves.
We discuss the former singularities in this section: we will see 
that these singularities do not dominate
in the asymptotic regime.
The dominating latter singularities are classified in the next section
in the two-dimensional setting.

Roots of the dispersion relation $\det(\Lambda-F(\omega))=0$ yield poles
for the functions $\tilde{a}_j$. Such poles in the upper half plane
correspond to eigenvalues of the linearized Vlasov operator, with
exponentially growing eigenmodes. Since we are interested in the
relaxation of a perturbation close to a stable stationary state of the
Vlasov equation, we assume that there are no such eigenmodes.  Poles
on the real axis correspond to purely oscillating eigenmodes.  In
order to study the decay of perturbations, we also assume that there
are no such modes.  There may be roots of the dispersion relation in
the lower half plane ${\mbox Im}(\omega)<0$ (in this case, they are rather
roots of the analytic continuation of the dispersion relation). These
are the usual ``Landau poles'', giving rise to exponential
damping. This damping may be an important feature of the dynamics at
intermediate time scales, especially if the pole is close to the real
axis~\cite{Weinberg94,Weinberg00,BOY10}, but the asymptotic regime is always
dominated by the singularities of the functions $F$'s and $G$'s, as
will become clear in the following sections.

\section{Singularities of $F$ and $G$ in two-dimensional setting}
\label{sec:singularities-FG}

We have seen that the functions $F$'s and $G$'s have no singularities
for ${\mbox Im}(\omega)>0$. The integrands of $F$'s and $G$'s are singular
for $\bJ$'s such that $\bm\cdot\bOmega(\bJ)-\omega$ vanishes,
and these singularities may yield branch points on the real $\omega$ axis.
Thus, to investigate the asymptotic behaviour of a perturbation
due to these branch points,
we turn now to the singularities on the real axis.
From now on and for simplicity, we restrict
ourselves to two-dimensional integrations over $\bJ=(J_1,J_2)$ in the
formulas (\ref{eq:F}) and (\ref{eq:G}) entering into the definitions
of $F$ and $G$.  We have in mind perturbations in spherically
symmetric self-gravitating systems, which fit into this
two-dimensional setting, as shown in section~\ref{sub:example}.
There would be no major obstacles to an analysis in
higher dimension, except the growing number and complexity of the
possible types of singularities.

\subsection{Spherically symmetric system}
\label{sub:example}
Although Vlasov-type equations appear in various settings, the main
situation we have in mind is the dynamics of a perturbation of a
spherical self-gravitating system. The dynamics in the potential
created by the stationary distribution is then integrable. It is
customary \cite{BinneyTremaine} to use as actions $J_{r}$ ("radial action"),
$L$ (modulus of the angular momentum) and $L_{z}$ (projection on the
$z$ axis of the angular momentum). The radial action $J_{r}$, energy $E$
of a particle and angular momentum $L$ are related by
\begin{equation}
    \label{eq:spherically-symmetric-system}
    J_{r}
    = \frac{1}{\pi} \int_{r_{min}}^{r_{max}} \sqrt{2E-2\Phi(r)-\frac{L^{2}}{r^{2}}}dr
\end{equation}
where $\Phi(r)$ is the potential created by the stationary state and
$r_{min}$ and $r_{max}$ are the minimum and maximum distance from the
origin reached by the particle. In this coordinate system,
the one-particle Hamiltonian depends only on the actions $J_{r}$ and $L$.

Let us assume that $f_{0}$ and $g$ depend on the actions
only through the one-particle Hamiltonian.
Then, $f_{0}(\bJ)$, $g(\bJ)$ and $c_{k}(\bm,\bJ)$
depend only on actions $J_{r}$ and $L$,
and hence the potentially singular integrands
in (\ref{eq:F}) and~(\ref{eq:G})
also depend only on $J_{r}$ and $L$.
The three-dimensional integrals are thus reduced to two-dimensional
integrals. The case of spherically symmetric
systems is then included in the abstract setting introduced in
section~\ref{sec:abstract_pbm}.

\subsection{An abstract problem}
\label{sec:abstract_pbm}
The abstract problem is to study the singularities of
the analytic continuation of $\varphi(z)$
defined for ${\mbox Im}(z)>0$ as integrals over a domain
$D\subset\mathbb{R}^2$:
\begin{equation}
    \varphi(z)
    =\int_{D\subset\mathbb{R}^2} \frac{\nume(\bJ)}{\denom(\bJ)-z}d\bJ
    \label{eq:varphi}
\end{equation}
with $\mu$ and $\nu$ real functions.
Functions $F$ and $G$, (\ref{eq:F}) and (\ref{eq:G}) respectively,
fit in this framework, with $\denom(\bJ)=\bm\cdot\bOmega(\bJ)$,
and by defining properly $\nume(\bJ)$.
We assume $\bm\neq\bzero$ since the
contributions from $\bm=\bzero$ in the definitions of $F$ and $G$ vanish.
We also assume that $\denom$ and $\nume$ are very regular:
although they are naturally defined over the domain $D\subset \mathbb{R}^2$,
they may be analytically continued over the complex $\bJ$ domain.
We further assume that their integrability properties are as good as needed.
Our goal is to study the singularities of the function
\begin{equation}
    \phi(x)=\lim_{y\to 0^+} \varphi(x+iy),~{ \mbox with}~x,y\in\mathbb{R}  .
    \label{eq:phi}
\end{equation}
This is the analytic continuation on the real axis of $\varphi(z)$, 
which is a priori defined for ${\mbox Im}(z)>0$. 

For $z$ real, the denominator in (\ref{eq:varphi}), $\denom(\bJ)-z$,
may vanish.  
We explain in~\ref{sec:continuation-details} why for a generic $z$ there
exists an analytic continuation of $\varphi$ in a neighborhood of $z$,
and hence there is no singularity for $\phi$.
From the computation in~\ref{sec:continuation-details}, special values
of $x_{0}=\denom(\bJ^\ast)$ corresponding to a singularity
for $\phi$ are easily identified; we classify 
the special points $\bJ^{\ast}$ and the associated
singularities for $\phi$ in the following subsections:
\begin{enumerate}
      \item Vertex singularity: 
    a point $\bJ^{\ast}$ at which
      the boundary of $D$ is not regular (section~\ref{sec:vertex})
      \item Tangent singularity: a point $\bJ^{\ast}$
      at which the level set $\denom(\bJ)=\denom(\bJ^{\ast})$
      is tangent to the boundary of $D$
      (section~\ref{sec:tangent}).
        \item Critical singularity: a critical point $\bJ^\ast$
      of the function $\denom(\bJ)$ inside $D$ 
      (section~\ref{sec:critical}).
\end{enumerate}
We will compute singularities of $\phi(x)$ around $x_{0}=\denom(\bJ^{\ast})$
by expanding $\denom(\bJ)$ around $\bJ^{\ast}$:
\begin{eqnarray}
    \denom(\bJ)
    &=& x_{0}
    + \denom_{1}(J_{1}-J_{1}^{\ast})
    + \denom_{2}(J_{2}-J_{2}^{\ast}) +\frac{1}{2} \denom_{11}(J_{1}-J_{1}^{\ast})^{2} \nonumber\\
    &+&\frac{1}{2} \denom_{22}(J_{2}-J_{2}^{\ast})^{2}
    + \denom_{12}(J_{1}-J_{1}^{\ast})(J_{2}-J_{2}^{\ast})
    + \cdots
    \label{eq:expansion-mu}
\end{eqnarray}
where, for instance, $\denom_{i}~(i=1,2)$ represents
$(\partial\denom/\partial J_{i})(\bJ^{\ast})$
and is not zero in general.
The above three types of singularities (i), (ii) and (iii)
appear at the point $\bJ^{\ast}$
where respectively (i) neither $\denom_{1}$ nor $\denom_{2}$ vanish,
(ii)  $\denom_{1}$ or $\denom_{2}$ vanishes, but not both,
and (iii) both $\denom_{1}$ and $\denom_{2}$ vanish.
The three types of singularities have therefore respectively
$0$, $1$ and $2$ analytic conditions for $\denom(\bJ)$, and 
$2$, $1$ and $0$ geometric conditions;
see Table~\ref{tab:vertex-tangent-critical}.
Considering that there are two variables $J_{1}$ and $J_{2}$,
these singularities appear generically.
Special treatments are devoted to the case of an infinite domain $D$
(section~\ref{sec:infinity}), and to a special situation which however does
arise generically for spherically symmetric self-gravitating systems
(section~\ref{sec:line}).

\begin{table}
    \centering
    \begin{tabular}{ccc}
        \hline
        {\bf Singularity} & {\bf Conditions for $\denom(\bJ)$}
        & {\bf Geometric conditions at $\bJ^{\ast}$} \\
        \hline
        Vertex & --- & On $\partial D$, non-regular point \\
        Tangent & $\denom_{1}=0$ or $\denom_{2}=0$ & On $\partial D$ \\
        Critical & $\denom_{1}=0$ and $\denom_{2}=0$ & --- \\
        \hline
    \end{tabular}
    \caption{Summary for the vertex, tangent and critical singularities.
      $\denom_{i}$ represents $(\partial\denom/\partial J_{i})(\bJ^{\ast})$
      for $i=1,2$, and is not zero in general.
      The symbol $\partial D$ denotes the boundary of $D$.
      The ``or'' in tangent singularity is exclusive.}
    \label{tab:vertex-tangent-critical}
\end{table}

\subsection{Vertex singularity}
\label{sec:vertex}
\begin{figure}
    \centering
    \includegraphics[width=10cm]{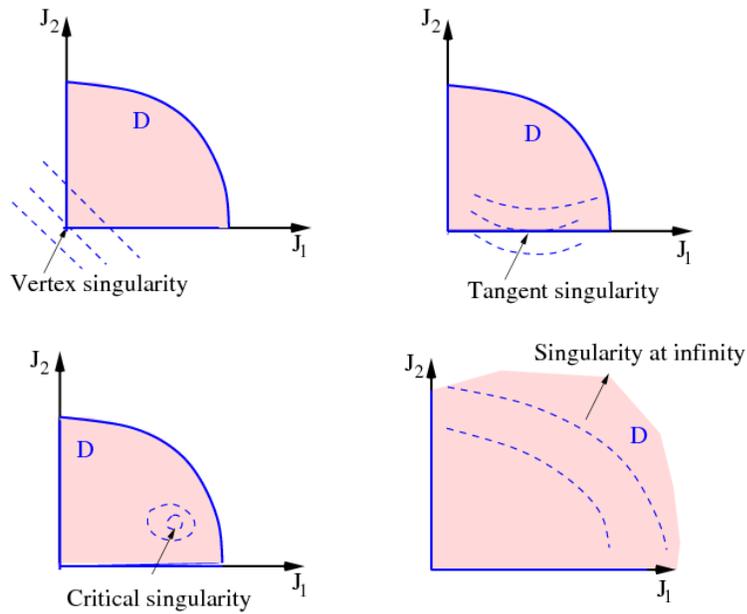}
    \caption{\label{fig:singu} Vertex (top left), tangent (top right), critical 
      (bottom left) singularities, and singularity at infinity (bottom right). In 
      all cases, the domain $D$ is shaded, and the dashed lines are level sets of 
the function $\denom(\bJ)$.}
\end{figure}

We consider first a singular point $\bJ^{\ast}$
on the boundary of the domain $D$, such that there is a jump in the
boundary's slope at $\bJ^{\ast}$. We refer to such a point as a
"vertex", see figure~\ref{fig:singu}. This type of 
singularity is always present in spherical self-gravitating systems
at zero radial action and zero angular momentum.

To investigate the leading singularity of $\phi(x)$
created by the vertex,
it is enough to keep only the leading order terms
in the expansions of $\denom(\bJ)$ and $\nume(\bJ)$.
We expand the function $\denom(\bJ)$ as
\begin{equation}
    \denom(\bJ)
    \simeq x_{0} + \denom_{1} (J_{1}-J_{1}^{\ast})
    + \denom_{2}(J_{2}-J_{2}^{\ast}) 
\end{equation}
where $x_{0}=\denom(\bJ^{\ast})$, and $\nume(\bJ)$ as
\begin{equation}
    \label{eq:g-leading}
    \nume(\bJ) \sim C (J_1-J_1^{\ast})^{a_1}(J_2-J_2^{\ast})^{a_2},
\end{equation}
where $a_{1}$ and $a_{2}$ are non-negative integers.
We assume that the first-order derivatives
$\denom_1$ and $\denom_2$ do not vanish,
which should be the generic case.
Here and in the following $C$ is a constant whose value plays no role
and may vary from line to line.

The singularity of $\phi(x)$ is then investigated on
 the following reduced expression for $\varphi(z)$
\begin{equation}
    \varphi(z)
    = C \int_{U_{\bJ^{\ast}}} \frac{(J_{1}-J_{1}^{\ast})^{a_{1}}(J_{2}-J_{2}^{\ast})^{a_{2}}}
    {\denom_{1}(J_{1}-J_{1}^{\ast})+\denom_{2}(J_{2}-J_{2}^{\ast})-(z-x_{0})}
    dJ_{1}dJ_{2},
\end{equation}
where $U_{\bJ^{\ast}}$ is a domain including a vertex at $\bJ=\bJ^{\ast}$.
Shifting and rescaling $J_{1}$ and $J_{2}$, we have
\begin{equation}
    \varphi(z)
    = C \int_{U_{0}} \frac{J_{1}^{a_{1}}J_{2}^{a_{2}}}{J_{1}+J_{2}-(z-x_{0})}
    dJ_{1}dJ_{2},
\end{equation}
where $U_{0}$ is a domain including the shifted vertex at $\bJ=0$.
Using new variables $u=J_{1}+J_{2}$ and $v=J_{1}-J_{2}$,
\begin{equation}
    \varphi(z)
    = \sum_{k,l\geq 0, k+l=a_{1}+a_{2}} c_{kl} \int_{U} 
    \frac{u^{k} v^{l}}{u-(z-x_{0})} du dv,
\end{equation}
where $c_{kl}$ is a constant and $U$ is a domain including a vertex at
$(u,v)=(0,0)$.
We assume that the line $u=0$ does not coincide with the boundary
of $U$, since this case results in the line singularity (see section~\ref{sec:line}).
As shown in \ref{sec:vertex-singularity-details},
the integral over $v$ yields
\begin{equation}
    \varphi(z) = C \int_{0}^{c} \frac{u^{1+a_{1}+a_{2}}}{u-(z-x_{0})} du,
\end{equation}
which fits in the framework (\ref{eq:singularity-computation-function1}).
Therefore, the singularity of $\phi(x)$ at $x=x_{0}$ is:
\begin{equation}
    \phi^{\rm sing}_{x_{0}}(x) = C_{1}(x-x_{0})^{a_{1}+a_{2}+1} \ln |x-x_{0}|
    + C_{2} (x-x_{0})^{a_{1}+a_{2}+1} H(x-x_{0}),
\end{equation}
where $H$ is the Heaviside step function.

\subsection{Tangent singularity}
\label{sec:tangent}
We consider now a regular point $\bJ^{\ast}$ 
on the boundary of the domain $D$
such that the level set $\denom(\bJ)=\denom(\bJ^{\ast})=x_{0}$
is tangent to the boundary of $D$ at $\bJ=\bJ^{\ast}$;
see figure~\ref{fig:singu}. 
The appearance of a tangent singularity requires that one derivative
of $\denom$ vanishes, at a point located on the
boundary of $D$.
Changing variables, we may assume that the boundary of $D$
is locally parametrized as $J_{2}=J_{2}^{\ast}$;
then the tangency condition reads
\begin{equation}
    \label{eq:tangent-condition}
    \left. \frac{\partial\denom}{\partial J_{1}} \right|_{\bJ=\bJ^{\ast}} = 0,
    \quad \bJ^{\ast}\in \partial D
\end{equation}
and at leading order we will use
\begin{equation}
    \denom(\bJ)
    \simeq x_{0} + \frac{1}{2}\denom_{11} (J_{1}-J_{1}^{\ast})^{2}
    + \denom_{2}(J_{2}-J_{2}^{\ast}) .
\end{equation}
Using also the leading order approximation (\ref{eq:g-leading}) for $\nume$,
and introducing these into (\ref{eq:varphi}), we obtain after
shifting and scaling $J_{1}$ and $J_{2}$
the reduced expression for $\varphi(z)$: 
\begin{equation}
    \varphi(z)
    = C \int_{U} \frac{J_{1}^{a_{1}}J_{2}^{a_{2}}}{J_{1}^{2}+J_{2}-(z-x_{0})}
    dJ_{1}dJ_{2}
\end{equation}
where the new tangent point is $\bJ^{\ast}=(0,0)$, and we have chosen
a subdomain $U$ close to $\bJ^{\ast}$.
Changing variables again and performing one integration,
the function $\varphi$ is further reduced to 
\begin{equation}
    \varphi(z)
    = C ( 1-(-1)^{1+a_{1}}) \int_{0}^{c} \frac{u^{1/2+a_{1}/2+a_{2}}}{u-(z-x_{0})} du,
\end{equation}
see \ref{sec:tangent-singularity-details} for details.
This form fits in the framework (\ref{eq:singularity-computation-function1});
hence the singularity of $\phi$ is:
\begin{equation}
    \phi^{\rm sing}_{x_{0}}(x) = \left\{
      \begin{array}{ll}
          C_{1} (x-x_{0})^{1/2+a_{1}/2+a_{2}} H(x-x_{0}) & \\
          \quad + C_{2} (x_{0}-x)^{1/2+a_{1}/2+a_{2}} H(x_{0}-x) 
          & a_{1} \mbox{:even} \\
          0 & a_{1} \mbox{: odd}.
      \end{array}
    \right.
\end{equation}

\subsection{Critical singularity}
\label{sec:critical}
We consider now
a point $\bJ^{\ast}$ belonging to the interior of $D$
such that the level set $\denom(\bJ)=\denom(\bJ^{\ast})=x_{0}$
is singular at $\bJ^{\ast}$;
this requires that both derivatives of $\denom$ vanish at $\bJ^{\ast}$.
In addition, we assume that this critical point is generic, thus the
Hessian at $\bJ^{\ast}$ has both eigenvalues different from $0$. We
also discard as non-generic the situation where the critical point
$\bJ^{\ast}$ belongs to the boundary of $D$; see an illustration on
figure~\ref{fig:singu}.  The critical point may be a local extremum or a
saddle point.  We study both cases in~\ref{sec:critical_appendix},
and summarize the results here.
Using the leading order approximation~(\ref{eq:g-leading}) for $\nume$,
changing variables and performing one integration,
the function $\varphi(z)$ is reduced to 
\begin{equation}
    \varphi(z) = C \int_{0}^{c} \frac{u^{(a_{1}+a_{2})/2}}{u-(z-x_{0})} du
\end{equation}
for a local extremum, and to
\begin{equation}
    \varphi(z) = C \int_{-c}^{c} \frac{u^{(a_{1}+a_{2})/2}\ln|u|}{u-(z-x_{0})} du
\end{equation}
for a saddle point. We note that these expressions are valid
for even $a_{1}$ and $a_{2}$, and no singularity appears otherwise.
These functions fit into (\ref{eq:singularity-computation-function1})
and (\ref{eq:singularity-computation-function2}) respectively,
and, for both extremum and saddle points, the singularity of $\phi$ is:
\begin{equation}
    \phi^{\rm sing}_{x_{0}}(x)
    = \left\{
      \begin{array}{ll}
          C_{1}(x-x_{0})^{(a_1+a_2)/2} \ln|x-x_{0}| & \\
            \quad +C_{2} (x-x_{0})^{(a_{1}+a_{2})/2} H(x-x_{0})
      & a_{1},a_{2}~\mbox{: even}\\
      0 & \mbox{otherwise.}
     \end{array}
    \right. 
\end{equation}

\subsection{Singularity "at infinity"}
\label{sec:infinity}

Assume $\denom(\bJ)$ tends to $0$ when $|\bJ|$ tends to infinity;
then $x_{0}=0$ is a singular point
for $\phi(x)$, see figure~\ref{fig:singu}.
This is a common situation for spherical self-gravitating systems.
To study this singularity, we assume that the domain $D$ is
$[0,+\infty)\times[0,+\infty)$, and
\begin{equation}
    \denom(\bJ) \sim (J_{1}+J_{2})^{-a} ~,~
    \nume(\bJ) \sim (J_{1}+J_{2})^{-b}.
\end{equation}
We assume that both $a$ and $b$ are integers
with $b>2$ to ensure convergence of the integrals in (\ref{eq:varphi}).

Changing variables to $u=J_{1}+J_{2}$, $v=J_{1}-J_{2}$,
and then from $u$ to 
$s=C'_{2}/u^{a}$, we obtain
\begin{eqnarray}
    \varphi(z)
    &=& C'_{1} \int_{\varepsilon}^{+\infty} du \int_{-u}^{u} dv
    \frac{u^{-b}}{C'_{2}u^{-a}-(z-x_{0})}  \\
    &=& C''_{1}\int_{0}^{c} \frac{s^{(b-2)/a-1}}{s-(z-x_{0})}ds.
\end{eqnarray}
The integral over $s$ fits in the framework
(\ref{eq:singularity-computation-function1}),
and the leading singularity of $\phi$ at $x_{0}$ is:
\begin{equation}
    \phi^{\rm sing}_{x_{0}}(x)
    = \left\{
      \begin{array}{ll}
          C_{1}(x-x_{0})^{(b-2)/a-1}\ln|x-x_{0}| \\
          ~~ + C_{2}(x-x_{0})^{(b-2)/a-1}H(x-x_{0})
          &  b-2\equiv 0~({\rm mod}~a) \\
          C_{1}(x-x_{0})^{(b-2)/a-1}H(x-x_{0}) \\
          ~~ + C_{2}(x_{0}-x)^{(b-2)/a-1}H(x_{0}-x)
          & b-2\not\equiv 0~({\rm mod}~a). \\
      \end{array}
    \right.
\end{equation}
We will find this kind of singularity in a model of self-gravitating systems, 
see section~\ref{sec:self-grav}.

\subsection{Line singularity}
\label{sec:line}

\begin{figure}
   \begin{minipage}[c]{.46\linewidth}
   \centerline{\includegraphics[width=7cm]{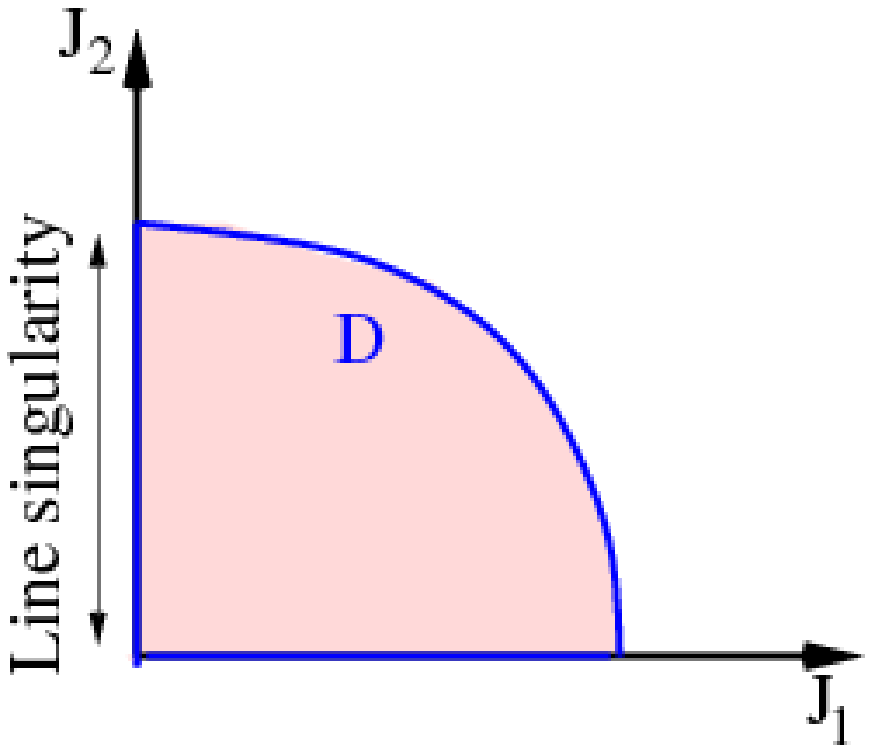}}
   \end{minipage} \hfill
   \begin{minipage}[c]{.46\linewidth}
      \centerline{\includegraphics[width=7cm]{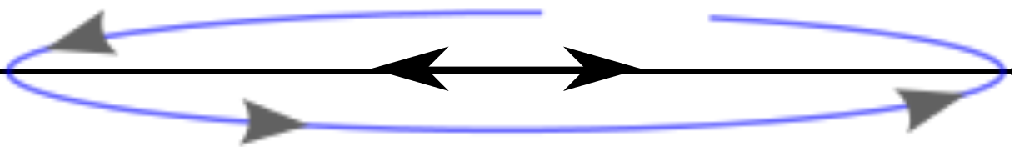}}
   \end{minipage}
   \caption{Left: schematic picture of a line singularity.
     Right: orbits of a particle in a spherical potential
     with $L=0$ (black line) and $L$ small (blue elongated orbit).}
   \label{fig:line-singularity}
\end{figure}

We consider now a case which seems at first sight very special, but
which does occur generically in spherical self-gravitating systems.
In a spherical potential, the orbit of a particle
with angular momentum $L=0$ is purely radial;
for small $L$, it is very elongated, and remains close 
to be purely radial. When the particle travels from one maximal radius
to the next along the elongated orbit, this corresponds to one radial period;
this also corresponds approximately to a half angular period,
see figure~\ref{fig:line-singularity}. Thus, on the $L=0$ line,
the radial frequency is exactly twice the angular frequency
$\bOmega_r=2\bOmega_{\theta}$; or equivalently $\bm\cdot\bOmega=0$,
with $\bm=(1,-2)$.

To study this singularity, we consider a domain $D$ as in figure~\ref{fig:line-singularity}: the
axis $J_{1}=0$ is a boundary. Suppose that
the function $\denom(\bJ)$ is constant on the $J_{2}$ axis as
\begin{equation}
    \label{eq:line_sing}
    \denom(0,J_{2}) = {\mbox const.} = x_0~,~\forall J_{2}\geq 0.
\end{equation}
We then expect a singularity of $\phi(x)$ at $x=x_{0}$.
Close to the singular line $J_{1}=0$, we expand $\denom(\bJ)$:
\begin{equation}
    \denom(\bJ) = x_{0} +W(J_{2}) J_{1}+\ldots
\end{equation}
Always having in mind self-gravitating systems, we assume that it is
possible to expand $\nume$ close to the $J_{2}$ axis as:
\begin{equation}
    \nume(\bJ) \sim h(J_{2}) J_{1}^{a_{1}}
\end{equation}
with a non-negative integer $a_{1}$. The function $\varphi$ around $z=x_{0}$ reads
\begin{eqnarray}
    \varphi(z)
    &=&  \int_{0}^{\varepsilon} dJ_{1} \int_{0}^{A} dJ_{2}
    \frac{h(J_{2})J_{1}^{a_{1}}}{W(J_{2})J_{1}-(z-x_{0})}  \\
    &=& \int_{0}^{A} dJ_{2} \frac{h(J_{2})}{W(J_{2})^{1+a_{1}}}
    \int_{0}^{\varepsilon/W(J_{2})} \frac{u^{a_{1}}du}{u-(z-x_{0})} .
\end{eqnarray}
The integral over $u$ fits in the framework
(\ref{eq:singularity-computation-function1});
 we conclude that the singularity of $\phi$ at $x_{0}$ is:
\begin{equation}
    \phi^{\rm sing}_{x_{0}}(x)
    = C_{1} (x-x_{0})^{a_{1}}\ln|x-x_{0}|
    +C_{2}(x-x_0)^{a_1}H(x-x_0).
\end{equation}
This computation relies on the fact that $W(J_{2})$ is well behaved
(for instance bounded away from $0$ and $\infty$).

\section{Asymptotic behaviour of a perturbation}
\label{sec:asymptotic}

The singularities in the functions $F$ and $G$ are passed on to the
coefficients $\tilde{a}_k(\omega)$.  Each singularity corresponds to a
decaying term in the inverse Laplace transform.  If
$\tilde{a}_k(\omega)$ had a finite number of singularities, the
asymptotic in time behaviour of $a_k(t)$ would be a sum of decaying
terms, each one corresponding to a singularity of
$\tilde{a}_k(\omega)$ (see Theorem 19 in \cite{Lighthill}). We expect that
this remains true in our situation, where $\tilde{a}_k(\omega)$
actually has an infinite number of singularities.

According to the analysis of 
section~\ref{sec:singularities-FG} the
singular part of $\tilde{a}_k(\omega)$
(or rather of its analytic continuation on the real axis)
close to a singular value $\omega_{0}$ is, for $\omega>\omega_{0}$:
\begin{eqnarray}
    \quad(\omega-\omega_0)^{n}\ln|\omega-\omega_0|\quad
    &\mbox{for}& ~  n\in \mathbb{Z}_{+},
    \nonumber \\
    {\mbox or}~(\omega-\omega_0)^{a} H(\omega-\omega_0)\quad
    &\mbox{for}& ~a\in \mathbb{R}\setminus\mathbb{Z}. \nonumber
\end{eqnarray}
Here $\mathbb{Z}_+$ is the set of non-negative integers. The behaviour is of course similar for $\omega<\omega_0$.
If a function $\phi$ presents any of the above singularities, we associate 
to it an asymptotic behaviour
of its inverse Laplace transform $\check{\phi}$, according to the following
rules~\cite{Lighthill}:
\begin{eqnarray}
    \phi^{\rm sing}_{\omega_{0}}(\omega)
    = C (\omega-\omega_{0})^{n}\ln |\omega-\omega_0|
    ~(n\in\mathbb{Z}_{+})
    &\to& \check{\phi}^{\rm as}(t) = C' \frac{e^{-i\omega_{0}t}}{t^{n+1}}~, \\
    \phi^{\rm sing}_{\omega_{0}}(\omega)
    = C (\omega-\omega_{0})^{a}H(\omega-\omega_0)
    ~(a\in\mathbb{R}\setminus\mathbb{Z})
    &\to& \check{\phi}^{\rm as}(t) = C' \frac{e^{-i\omega_{0}t}}{t^{a+1}}~. 
\end{eqnarray}
Singularities and associated asymptotic behaviour are summarized
in Table~\ref{tab:singularity-alternate}.
Thus, to extract the asymptotic behaviour of $a_k(t)$, one needs to
\begin{enumerate}
\item Enumerate the singularities appearing in $\tilde{a}_k(\omega)$
\item Keep the strongest singularity, corresponding to the slowest decay
\item Check if the symmetries of the problem induce a special cancellation. 
\end{enumerate}
Unless a cancellation occurs, this strongest singularity yields the
decay exponent and asymptotic frequency of $a_k(t)$. We illustrate
this strategy on several examples in the following sections, 
including the case of a cancellation; a careful 
reexamination of the singularity is then needed.

\begin{table}[h]
    \centering
    \begin{tabular}{ll|cll|l}
        \hline
        {\bf Type}
        &
        & {\bf Singularity}
        & {\bf Exponent $\alpha$} 
        & {\bf Damping} 
        & {\bf Sign} \\
        \hline
        Vertex
        &
        & (\ref{eq:singularity-function1a})
        & $1+a_{1}+a_{2}$ 
        & $e^{-ix_{0}t}t^{-(2+a_{1}+a_{2})}$ 
        & $(-1)^{1+\alpha}$ \\
        \hline
        Tangent
        & ($a_{1}$ : even)
        & (\ref{eq:singularity-function1b})
        & $1/2+a_{1}/2+a_{2}$ 
        & $e^{-ix_{0}t}t^{-(3/2+a_{1}/2+a_{2})}$
        & ----- \\
        & ($a_{1}$ : odd)
        & N/A
        & N/A
        & N/A\\
        \hline
        Critical & & & & \\
        extremum 
        & ($a_{1},a_{2}$ : even)
        & (\ref{eq:singularity-function1a})
        & $(a_{1}+a_{2})/2$ 
        & $e^{-ix_{0}t}t^{-(1+(a_{1}+a_{2})/2)}$ 
        & $(-1)^{1+\alpha}$ \\
        saddle
        & ($a_{1},a_{2}$ : even)
        & (\ref{eq:singularity-function2})
        & $(a_{1}+a_{2})/2$ 
        & $e^{-ix_{0}t}t^{-(1+(a_{1}+a_{2})/2)}$
        & $(-1)^{\alpha}$ \\
        & (otherwise)
        & N/A
        & N/A
        & N/A \\
        \hline
        Infinity
        & ($(b-2)/a\in\mathbb{Z}$)
        & (\ref{eq:singularity-function1a})
        & $(b-2)/a-1$ 
        & $e^{-ix_{0}t}t^{-(b-2)/a}$
        & $(-1)^{1+\alpha}$ \\
        & ($(b-2)/a\notin\mathbb{Z}$)
        & (\ref{eq:singularity-function1b})
        & $(b-2)/a-1$ 
        & $e^{-ix_{0}t}t^{-(b-2)/a}$
        & ----- \\
        \hline
        Line
        &
        & (\ref{eq:singularity-function1a})
        & $a_{1}$
        & $e^{-ix_{0}t}t^{-(1+a_{1})}$
        & $(-1)^{1+\alpha}$ \\
        \hline
    \end{tabular}
    \caption{
      Table of singularities of the function $\phi(x)$ 
      defined by (\ref{eq:varphi}) and (\ref{eq:phi}),
      with their associated exponent $\alpha$
      (see section \ref{sec:singularities-FG} and 
      \ref{sec:main_computation}), and asymptotic dampings.
      The singularity is produced at $\bJ^{\ast}=(J_{1}^{\ast},J_{2}^{\ast})$, and
      $x_{0}=\denom(J_{1}^{\ast},J_{2}^{\ast})$.      
      The leading order of the numerator is
      $\nume(\bJ)\simeq (J_{1}-J_{1}^{\ast})^{a_{1}}(J_{2}-J_{2}^{\ast})^{a_{2}}$
      for vertex, tangent and critical singularities,
      $\nume(\bJ)\simeq (J_{1}+J_{2})^{-b}$ for a singularity at infinity,
      and $\nume(\bJ)\simeq h(J_{2})J_{1}^{a_{1}}$ for a line singularity.
      We assume $\denom(\bJ)\simeq (J_{1}+J_{2})^{-a}$
      for a singularity at infinity.
      $a_{1}$ and $a_{2}$ are non-negative integers,
      $a$ is an integer and $b$ is an integer satisfying $b>2$.
      N/A means that no singularity appears in $\phi(x)$.
      This table shows the leading singularity for each type.
      As shown in sections \ref{sec:numerical-tests}
      and \ref{sec:self-grav}, a special cancellation 
      for the leading term may happen between modes $\bm$ and $-\bm$. 
      The column Sign represents the relative sign
      between the singular parts of modes $\bm$ and$-\bm$
      for $x_{0}=0$. No cancellation is expected for $x_{0}\neq 0$
      irrespective of the relative sign.
      A bar means no simple relation between the two modes.
      Note that the relative sign may also depend on
      the numerator $\nu(\bJ)$ of (\ref{eq:varphi}) if it depends on $\bm$,
      but this dependence is ignored in this table.
      }
    \label{tab:singularity-alternate}
\end{table}

A remark is in order: we study here the asymptotic behaviour of the
$a_k(t)$, which are the coefficients of the density and potential
perturbation in the expansions~(\ref{eq:rho1_exp})
and~(\ref{eq:phi1_exp}). One might expect that the slowest decay for
these coefficients yields the asymptotic decay of the density or
potential perturbation at a given point in space. This is not
necessarily true, as we have no information on the rate at which $a_k$
reaches its asymptotic regime, and thus have no control on the 
series~(\ref{eq:rho1_exp}) and~(\ref{eq:phi1_exp}).

\section{Preparation for numerical tests}
\label{sec:preparation-numerics}

\subsection{On numerical tests}
The theory developed in this paper should describe the asymptotic
behaviour of a perturbation evolving according to the linearized Vlasov
equation. Two questions arise: \\
i) Since the above computations are
formal, one may wonder if the theory is correct for the evolution of a
linearized Vlasov equation.\\
ii) Even if the theory is correct for the linearized Vlasov equation,
it is not clear that it describes
the asymptotic behaviour of the fully non-linear Vlasov equation.\\
To answer these questions, direct numerical simulations of the Vlasov
equation are required. Such a check has been done in a one-dimensional setting
using $N$-body simulations~\cite{BOY11};
however, the memory, accuracy and time frame required to
perform similar simulations in two or more spatial dimensions make the task
challenging.  Instead, we will illustrate the various
aspects of the theory on much simpler linear advection equations
in section~\ref{sec:numerical-tests},
and apply it to make definite predictions on
a spherically symmetric self-gravitating system
in section~\ref{sec:self-grav}.

\subsection{Advection equation and singularity analysis}
\label{sec:advection}

To illustrate the theory, we consider an advection equation
\begin{equation}
    \label{eq:advection}
    \partial_{t} f + \bOmega(\bJ)\cdot\nabla_{\bt}f = 0.
\end{equation}
The exact solution is 
\begin{equation}
    \label{eq:advection-exact-solution}
    f(\bt,\bJ,t)
    = f(\bt-\bOmega(\bJ)t,\bJ,t=0).
\end{equation}
We consider the temporal evolution of the expected value of $A(\bt)$
defined by
\begin{equation}
    \left\langle A \right\rangle(t)
    = \int A(\bt) f(\bt,\bJ,t) d\bt d\bJ,
\end{equation}
where $f$ is governed by the advection equation (\ref{eq:advection})
and we assumed that $A$ is a function of $\bt$ only.
From the Fourier-Laplace transform
of the advection equation (\ref{eq:advection}),
the Laplace transform of $\left\langle A \right\rangle(t)$
is given by
\begin{equation}
    \label{eq:fourier-laplace-A}
    \widetilde{\left\langle A \right\rangle}(\omega)
    = \sum_{\bm} \int A(\bt) e^{i\bm\cdot\bt} d\bt
    \int \frac{g(\bm,\bJ)}{\bm\cdot\bOmega(\bJ)-\omega} d\bJ,
\end{equation}
where $ig(\bm,\bJ)$ is the Fourier transform of $f(\bt,\bJ,t=0)$.
The function
\begin{equation}
    \label{eq:toy-varphi}
    \varphi(\omega;\bm)
    = \int \frac{g(\bm,\bJ)}{\bm\cdot\bOmega(\bJ)-\omega}d\bJ
\end{equation}
fits in the framework (\ref{eq:varphi}),
thus the asymptotic damping of the expected value
$\left\langle A \right\rangle(t)$
is predicted by the theory presented in
sections \ref{sec:singularities-FG} and \ref{sec:asymptotic}.

The theoretically predicted asymptotic damping
can be checked without computing the temporal evolution of $f$ numerically,
since we have the exact solution (\ref{eq:advection-exact-solution})
to the advection equation, which gives
\begin{equation}
    \label{eq:exact-expected-value}
    \left\langle A \right\rangle(t)
    = \sum_{\bm} \int A(\bt) e^{i\bm\cdot\bt} d\bt
    \int i g(\bm,\bJ) e^{-i\bm\cdot\bOmega(\bJ)t} d\bJ.
\end{equation}
Integrals over $\bt$ can be performed analytically,
and hence our numerical task is to perform integrals over $\bJ$.
Consequently, we can reduce the $2d$-dimensional problem
(\ref{eq:advection}) to $d$-dimensional integrals over $\bJ$.
For $2$-dimensional cases, numerical integrations are
possible with good accuracy.

\section{Numerical tests }
\label{sec:numerical-tests}
We set the initial distribution as
\begin{equation}
    \label{eq:toy-initial}
    f(\bt,\bJ,t=0) =
    J_{1}^{h_{1}}J_{2}^{h_{2}}
    (J_{1}-J_{1}^{\ast})^{a_{1}} (J_{2}-J_{2}^{\ast})^{a_{2}}
    e^{-(J_{1}^{2}+J_{2}^{2})/2} (1+\cos\theta_{1}\cos\theta_{2}),
\end{equation}
where the unity is to keep $f$ positive,
but is not important since the mode $(0,0)$ is temporally invariant.
We therefore focus on the four modes
$\bm=(1,1),(1,-1),(-1,1)$ and $(-1,-1)$.
We assume that the system is defined on the domain 
$(J_{1},J_{2})\in [0,+\infty)^{2}$.
The factor $(J_{1}-J_{1}^{\ast})^{a_{1}}(J_{2}-J_{2}^{\ast})^{a_{2}}$
corresponds to the leading estimation of $\nu(J)$ (\ref{eq:g-leading}), 
and the factor $J_{1}^{h_{1}}J_{2}^{h_{2}}$ will be used
to select the singularities we want to illustrate.
The function $g$ is the same for the four modes and,
remembering $ig$ is the Fourier transform of the initial perturbation,
\begin{equation}
    \label{eq:toy-g}
    ig(\bm,\bJ)
    = \frac{1}{4} J_{1}^{h_{1}}J_{2}^{h_{2}}
    (J_{1}-J_{1}^{\ast})^{a_{1}} (J_{2}-J_{2}^{\ast})^{a_{2}}
    e^{-(J_{1}^{2}+J_{2}^{2})/2},
    \quad \bm=(\pm 1,\pm 1).
\end{equation}
We introduce four observables:
\begin{equation}
    \label{eq:four-observables}
    A_{1} = \cos(\theta_{1}+\theta_{2}),~ 
    A_{2} = \sin(\theta_{1}+\theta_{2}),~
    A_{3} = \cos(\theta_{1}-\theta_{2}),~
    A_{4} = \sin(\theta_{1}-\theta_{2}),
\end{equation}
to pick up singularities appearing
in the modes $(1,1)$ and $(-1,-1)$ by $A_{1}$ and $A_{2}$,
and in the modes $(1,-1)$ and $(-1,1)$ by $A_{3}$ and $A_{4}$.

We now report detailed theoretical predictions and numerical illustrations
for vertex, tangent and critical singularities
in sections \ref{sec:vertex-singularity-numerics},
\ref{sec:tangent-singularity-numerics}
and \ref{sec:critical-singularity-numerics} respectively.
These include cases with cancellations between two modes.
The line singularity is discussed in section~\ref{sec:toy_model}
on a toy model, which also has the vertex and tangent singularities.  
We discuss the dominant singularity for different modes, again 
taking into account cancellations, and check the predictions against 
numerical computations. 
The singularity at infinity will be demonstrated in section~\ref{sec:self-grav}
on a spherical self-gravitating model.

In this section, all the numerical integrations for the exact expected values
are performed by introducing a cutoff at $J_{1}=J_{2}=10$
and slicing the interval $[0,10]$ in $2^{12}$ bins.

\subsection{Vertex singularity}
\label{sec:vertex-singularity-numerics}
We set the frequencies as
\begin{equation}
    \Omega_{1}(\bJ) = J_{1},
    \quad
    \Omega_{2}(\bJ) = J_{2},
\end{equation}
which yields a vertex singularity at the origin,
with $(J_{1}^{\ast},J_{2}^{\ast})=(0,0)$;
the associated frequency is $\omega_{0}=\bm\cdot\bOmega(\bJ^{\ast})=0$.
We compute $\left\langle A_{i}\right\rangle(t)$
for $a_{1}=0,1$ and $2$ with fixed $a_{2}=0$.
There are no singularities of other types,
thus we may use $h_{1}=h_{2}=0$.
From Table~\ref{tab:singularity-alternate},
we predict dampings and cancellations as shown in
Table~\ref{tab:vertex-singularity}.
Cancellations occur for $A_{1}$ and $A_{3}$ with an odd $a_{1}$,
since the cosine observables pick up the sum of the modes
$\bm$ and $-\bm$; for instance:
\begin{equation}
    \label{eq:toy-line-cos}
    A_{1} = \cos(\theta_{1}+\theta_{2})
    \quad \Longrightarrow \quad
    \widetilde{\left\langle A_{1} \right\rangle}(\omega)
    \propto \varphi(\omega;(1,1)) + \varphi(\omega;(-1,-1)),
\end{equation}
and $\varphi(\omega;(-1,-1))=(-1)^{a_{1}+a_{2}}\varphi(\omega;(1,1))$ holds
for the vertex singularity.
Similarly, cancellations occur for the sine observables
$A_{2}$ and $A_{4}$ with an even $a_{1}$ since, for instance,
\begin{equation}
    \label{eq:toy-line-sin}
    A_{2} = \sin(\theta_{1}+\theta_{2})
    \quad \Longrightarrow \quad
    \widetilde{\left\langle A_{2} \right\rangle}(\omega)
    \propto \varphi(\omega;(1,1)) - \varphi(\omega;(-1,-1)).
\end{equation}
We note that the cancellations occur for the leading singularities;
higher-order singularities may survive in general.
However, in this case, higher-order singularities
coming from the expansion of $\exp(-(J_{1}^{2}+J_{2}^{2})/2)$ 
do not change the parity of $a_{1}+a_{2}$,
since expanded terms consist of $J_{1}^{2}$ and $J_{2}^{2}$.
Thus, all orders are cancelled when the leading order is cancelled.

\begin{figure}
    \centering
    \includegraphics[width=6cm]{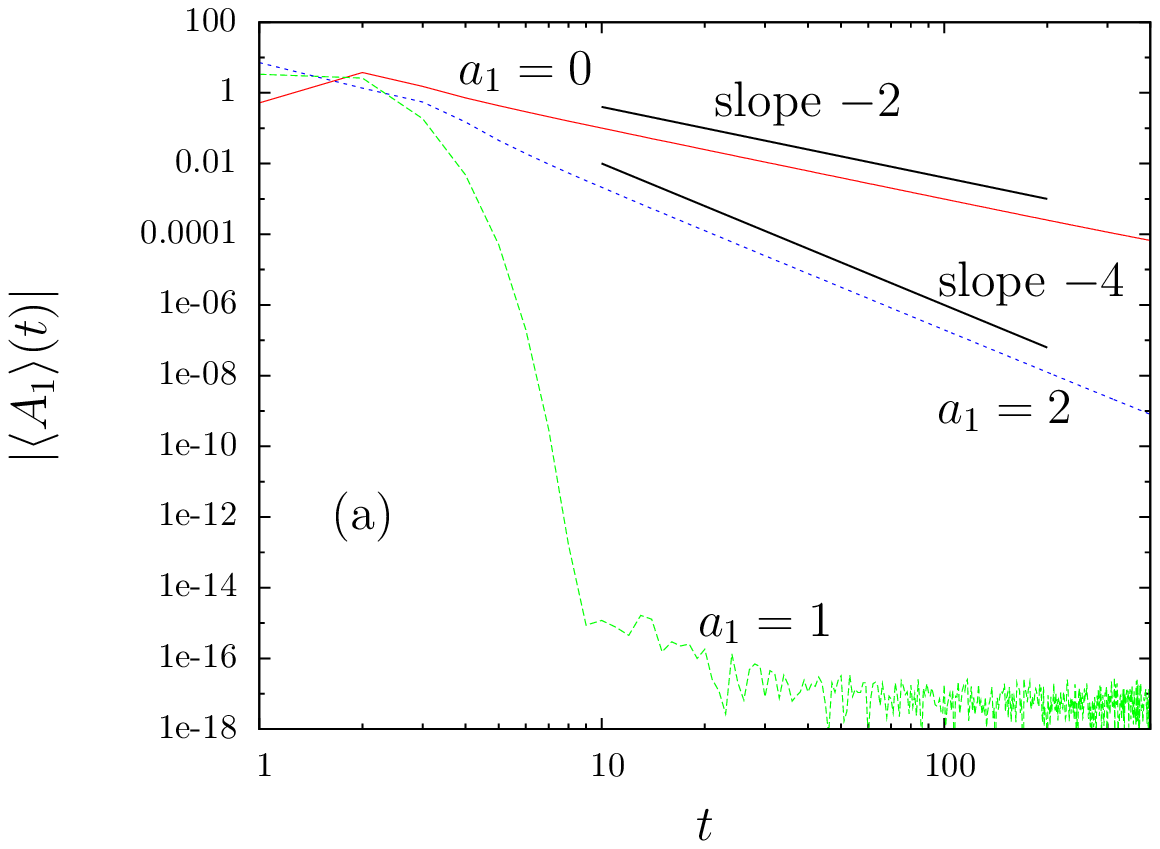}
    \includegraphics[width=6cm]{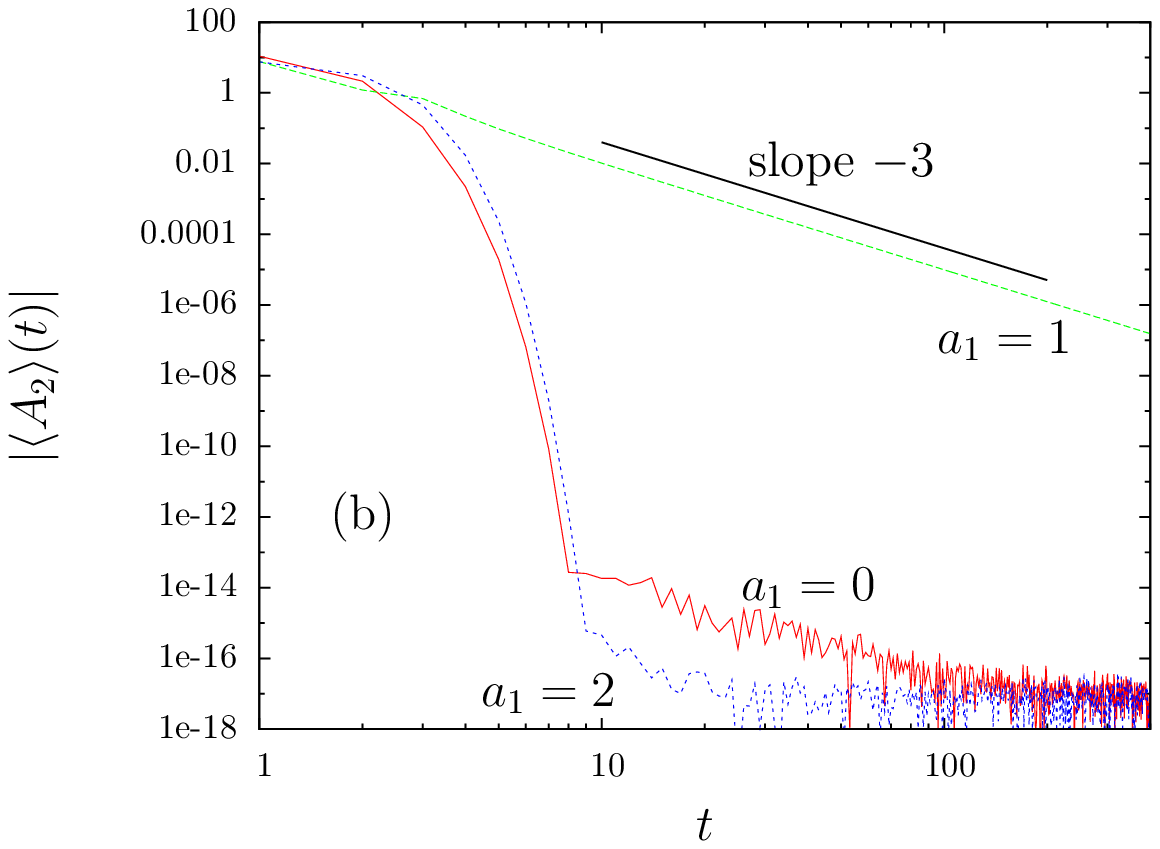}
    \caption{Dampings for the vertex singularity.
      Observables are (a) $A_{1}$ and (b) $A_{2}$.
      The parameter $a_{2}$ is fixed as $0$.
      Observables $A_{3}$ and $A_{4}$ give qualitatively
      similar panels to $A_{1}$ and $A_{2}$ respectively.}
    \label{fig:vertex-singularity}
\end{figure}

The above prediction is confirmed by direct numerical integration
of (\ref{eq:exact-expected-value}) as shown in
figure~\ref{fig:vertex-singularity}.
We remark that dampings without oscillation reflect the 
zero frequency $\omega_{0}=0$.
We have also checked that no cancellation occurs when we replace
$\Omega_{1}$ by $\Omega_{1}(\bJ)=J_{1}+1$,
since $\omega_{0}=1\neq 0$ (figure not shown).

\begin{table}
    \centering
    \begin{tabular}{c|ccc}
        \hline
        $(a_{1},a_{2})$
        & $(0,0)$
        & $(1,0)$
        & $(2,0)$ \\
        \hline
        $\left\langle A_{1}\right\rangle(t)$
        & $t^{-2}$
        & C
        & $t^{-4}$ \\
        $\left\langle A_{2}\right\rangle(t)$
        & C
        & $t^{-3}$
        & C\\
        $\left\langle A_{3}\right\rangle(t)$
        & $t^{-2}$ 
        & C
        & $t^{-4}$ \\
        $\left\langle A_{4}\right\rangle(t)$
        & C
        & $t^{-3}$
        & C\\
        \hline
    \end{tabular}
    \caption{Dampings and cancellations for the vertex singularity.
    ``C'' stands for "cancellation".}
    \label{tab:vertex-singularity}
\end{table}

\subsection{Tangent singularity}
\label{sec:tangent-singularity-numerics}
We set the frequencies as
\begin{equation}
    \Omega_{1}(\bJ) = (J_{1}-1)^{2}, \quad
    \Omega_{2}(\bJ) = J_{2}.
\end{equation}
There are a vertex singularity at $(J_{1},J_{2})=(0,0)$
and a tangent singularity at $(J_{1},J_{2})=(1,0)$.
We are interested in the tangent singularity,
and hence we set $(J_{1}^{\ast},J_{2}^{\ast})=(1,0)$
which gives the frequency $\omega_{0}=0$.
We choose $a_{1}$ and $a_{2}$ from the set $\{0,1,2\}$,
and use $h_{1}=2$ and $h_{2}=0$ to hide the vertex singularity:
the slowest decay coming from the vertex singularity is then $t^{-4}$.
The predicted dampings and cancellations are shown
in Table~\ref{tab:tangent-singularity}.
An explanation is needed for the case $(a_{1},a_{2})=(1,0)$:
the leading order does not give any singularity, since $a_{1}$ is odd;
however, expanding $ig(\bm,\bJ)$ around $(J_{1}^{\ast},J_{2}^{\ast})=(1,0)$,
the second leading order gives $a_{1}=2$;
thus the damping must be $t^{-2.5}$, as in the case $(a_{1},a_{2})=(2,0)$.
We remark that no cancellation occurs for the tangent singularities.
These predictions have been checked numerically,
though no figures are reported.

\begin{table}[h]
    \centering
    \begin{tabular}{c|ccccc}
        \hline
        $(a_{1},a_{2})$
        & $(0,0)$
        & $(0,1)$
        & $(0,2)$
        & $(1,0)$ 
        & $(2,0)$ \\
        \hline
        $\left\langle A_{i}\right\rangle(t)$
        & $t^{-1.5}$
        & $t^{-2.5}$
        & $t^{-3.5}$
        & $t^{-2.5}$
        & $t^{-2.5}$ \\
        \hline
    \end{tabular}
    \caption{Dampings and cancellations for the tangent singularity.}
    \label{tab:tangent-singularity}
\end{table}

\subsection{Critical singularity}
\label{sec:critical-singularity-numerics}
We set the frequencies as
\begin{equation}
    \Omega_{1}(\bJ) = (J_{1}-1)^{2},
    \quad
    \Omega_{2}(\bJ) = (J_{2}-1)^{2}.
\end{equation}
Singularities are: a vertex singularity at $(J_{1},J_{2})=(0,0)$,
tangent singularities at $(J_{1},J_{2})=(1,0)$ and $(0,1)$,
and a critical singularity at $(J_{1},J_{2})=(1,1)$.
We are now interested in the critical singularity,
and hence we set $(J_{1}^{\ast},J_{2}^{\ast})=(1,1)$
giving the frequency $\omega_{0}=0$.
We choose $a_{1}$ from the set $\{0,1,2\}$,
and use $h_{1}=h_{2}=2$ to hide the vertex and tangent singularities
as done in section~\ref{sec:tangent-singularity-numerics}.

Cancellations between the modes $\bm$ and $-\bm$ are as follows.
We first remark that no leading singularity appears 
for $(a_{1},a_{2})=(1,0)$ since $a_{1}$ is odd.
In this case, as discussed in the case of the tangent singularity,
the second leading singularity gives
the same singularity as $(a_{1},a_{2})=(2,0)$,
thus we concentrate on $(a_{1},a_{2})=(0,0)$ and $(2,0)$.
The second remark is that the critical singularity at
$(J_{1},J_{2})=(1,1)$ behaves as an extremum singularity
for $A_{1}$ and $A_{2}$,
and as a saddle singularity for $A_{3}$ and $A_{4}$.
Referring to Table~\ref{tab:singularity-alternate}
and the above second remark,
the leading singularities in $A_{1}$ and $A_{4}$
are cancelled for $(a_{1},a_{2})=(0,0)$,
and in $A_{2}$ and $A_{3}$ for $(a_{1},a_{2})=(2,0)$.
In the latter case, 
the next leading singularities come from $(a_{1},a_{2})=(4,0)$
for instance, and $t^{-3}$ dampings are predicted.
These discussions are summarized in Table \ref{tab:critical-singularity}.

We explain the whole cancellation in
$\left\langle A_{4}\right\rangle(t)$ for $(a_{1},a_{2})=(0,0)$.
The exact solution of $\langle A_{4}\rangle(t)$ is proportional to
\begin{equation}
    \left\langle A_{4}\right\rangle(t)
    = C
    \int g(J_{1},J_{2})\sin((\Omega_{1}(J_{1},J_{2})-\Omega_{2}(J_{1},J_{2}))t) d\bJ.
\end{equation}
Exchanging the dummy variables $J_{1}$ and $J_{2}$,
and using invariance of $g$ under this exchange, 
we have $\left\langle A_{4}\right\rangle(t)=0$.
From the view point of singularity analysis,
using notations of \ref{sec:critical_appendix}
and \ref{sec:main_computation},
the invariance of $g$ with respect to the exchange
between $J_{1}$ and $J_{2}$ implies that the numerator
of the integrand in (\ref{eq:singularity-computation-function2})
is of the type $u^{\alpha}\ln|u|$ with even $\alpha$,
since $u=J_{1}^{2}-J_{2}^{2}$.
The evenness of $\alpha$ gives the same sign
for the two modes $\bm$ and $-\bm$ in the saddle singularity,
hence they cancel each other in the sine observable $A_{4}$: this is
of course compatible with the exact result $\langle A_4 \rangle (t)=0$.

\begin{table}
    \centering
    \begin{tabular}{c|ccccc}
        \hline
        $(a_{1},a_{2})$
        & $(0,0)$
        & $(1,0)$ 
        & $(2,0)$ \\
        \hline
        $\left\langle A_{1}\right\rangle(t)$
        & $t^{-2}$(C)
        & $t^{-2}$(N)
        & $t^{-2}$ \\
        $\left\langle A_{2}\right\rangle(t)$
        & $t^{-1}$
        & $t^{-3}$(N)
        & $t^{-3}$(C) \\
        $\left\langle A_{3}\right\rangle(t)$
        & $t^{-1}$
        & $t^{-3}$(N)
        & $t^{-3}$(C) \\
        $\left\langle A_{4}\right\rangle(t)$
        & C
        & $t^{-2}$(N)
        & $t^{-2}$ \\
        \hline
    \end{tabular}
    \caption{Dampings and cancellations for the critical singularity.
    "C'' means that there is a cancellation at leading order,
    and ``N'' that the leading order singularity does not exist.
    Dampings associated with the next leading singularities
    are observed when ``C'' or ``N'' appears.
    In $\left\langle A_{4}\right\rangle(t)$ for $(a_{1},a_{2})=(0,0)$,
    all the singularities at any orders are cancelled
    due to a special symmetry of $g(\bm,\bJ)$.}
    \label{tab:critical-singularity}
\end{table}

\begin{figure}
    \centering
    \includegraphics[width=6cm]{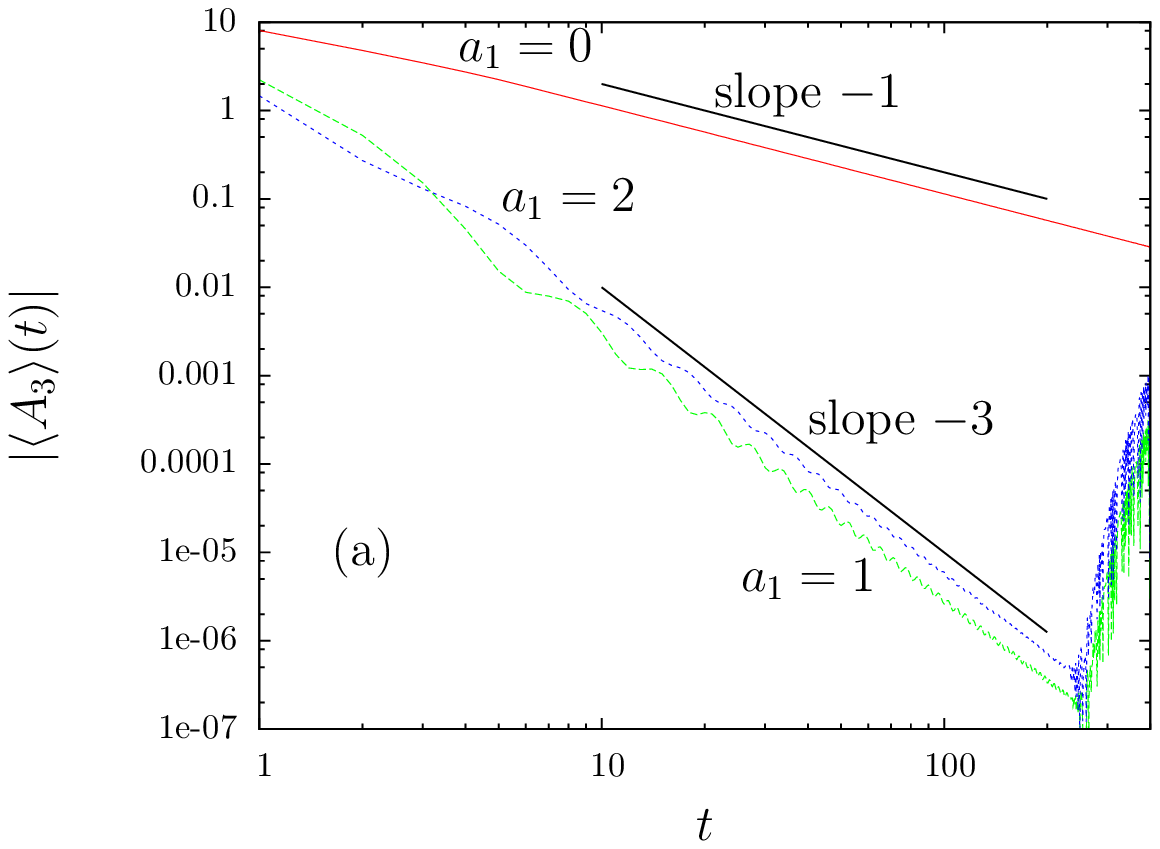}
    \includegraphics[width=6cm]{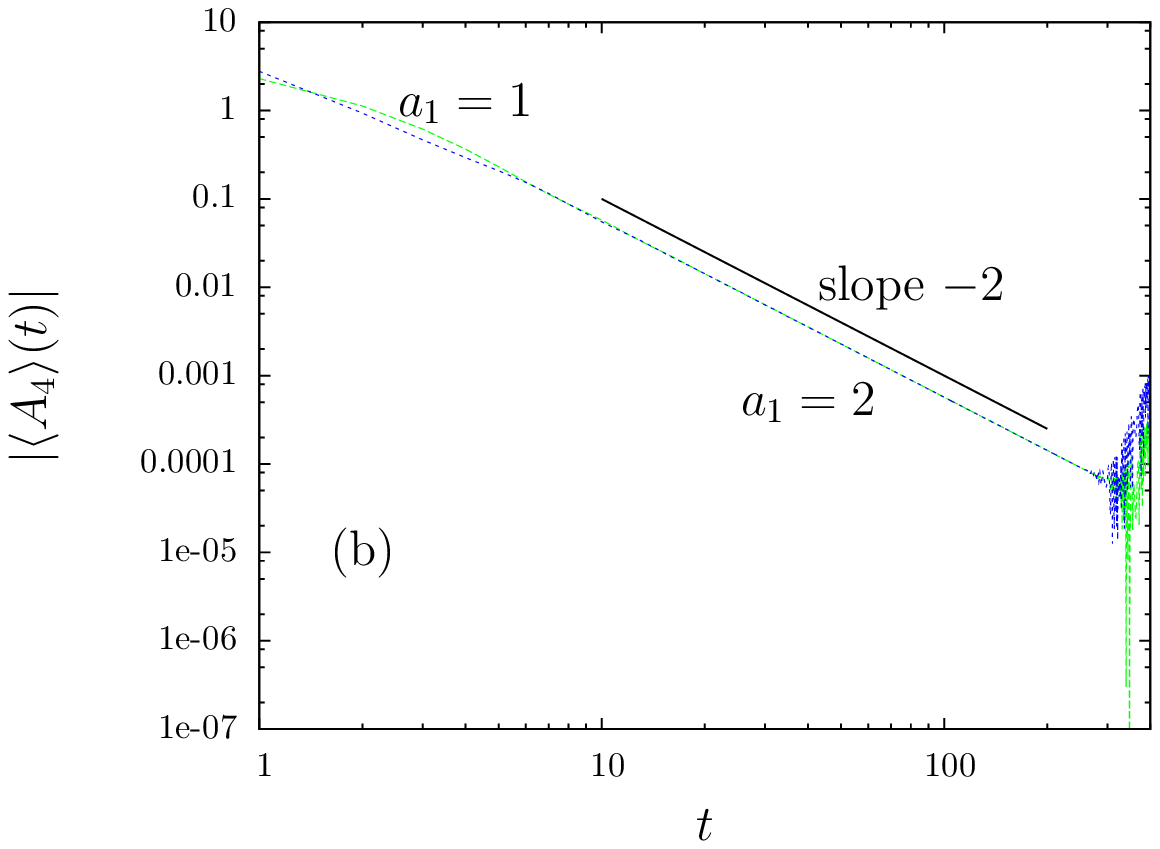}
    \caption{Dampings for the critical singularity.
      Observables are (a) $A_{3}$ and (b) $A_{4}$.
      The parameter $a_{2}$ is fixed as $0$.
      The curves for $a_{1}=1$ and $2$ in (b) are almost superposed.
      The curve of $a_{1}=0$ in (b) is out of range
      due to a special cancellation. Except for this special cancellation,
      $A_{1}$ and $A_{2}$ give qualitatively similar panels
      to $A_{4}$ and $A_{3}$ respectively.}
    \label{fig:critical-singularity}
\end{figure}

\subsection{Toy model}
\label{sec:toy_model}
We set the frequencies as
\begin{equation}
    \label{eq:toy-Omega}
    \Omega_{1}(\bJ) = -J_{1} - \frac{J_{1}^{2}}{2} - 2J_{2}, \quad
    \Omega_{2}(\bJ) = -2J_{1}+2J_{2}.
\end{equation}
This toy model (\ref{eq:toy-Omega}) has three types of singularities:
\begin{itemize}
      \item Vertex singularity
    at $(J_{1}^{\ast},J_{2}^{\ast})=(0,0)$
    for the modes $(1,-1)$ and $(-1,1)$.
      \item Tangent singularity
    at $(J_{1}^{\ast},J_{2}^{\ast})=(1,0)$
    for the modes $(1,-1)$ and $(-1,1)$.
      \item Line singularity
    on the $J_{2}$ axis
    for the modes $(1,1)$ and $(-1,-1)$.
\end{itemize}
No critical singularity and no singularity at infinity appear.
We use distribution (\ref{eq:toy-initial}),
with $h_{1}=h_{2}=a_{1}=a_{2}=0$, that is
\begin{equation}
    ig(\bm,\bJ) = e^{-(J_{1}^{2}+J_{2}^{2})/2}.
\end{equation}

\subsubsection{Cancellation and higher-order contribution}
Three different types of singularities appear for the
$A_i$ observables, the strongest of which is a line singularity, see
Table~\ref{tab:toy-theory}. Concentrating on $A_{1}$ and $A_{2}$, we
show an example of cancellation for this line singularity. 

The relative sign between $\varphi(\omega;\bm)$
and $\varphi(\omega;-\bm)$ is $(-1)^{1+a_{1}}$
for the line singularity.
Thus, the leading singularity corresponding to $a_{1}=0$ 
is cancelled for $A_{1}$ but not for $A_{2}$.

Higher-order singularities 
may come from the expansion of $g$ around $J_{1}=0$
and correspond to $a_{1}\geq 1$.
However, since $g$ is even in $J_{1}$,
the same cancellation will take place at all orders.
Consequently, there is no higher-order contribution from $g$.

Nevertheless, a higher-order contribution comes from 
$\bm\cdot\bOmega(\bJ)$. For the mode $\bm=(1,1)$,
the function $\varphi(\omega;\bm)$ becomes
\begin{equation}
    \varphi(\omega;(1,1))
    \simeq \frac{1}{4i} \int_{0}^{\epsilon}
    \frac{dJ_{1}}{-3J_{1}-J_{1}^{2}/2-\omega} \int_{0}^{C} dJ_{2}
    = \frac{C}{4i} \int_{0}^{\epsilon}
    \frac{dJ_{1}}{-3J_{1}-J_{1}^{2}/2-\omega}
\end{equation}
keeping the leading order of $g(\bm,\bJ)$, which is constant.
Changing variable from $J_{1}$ to $u=3J_{1}+J_{1}^{2}/2$, we have
\begin{eqnarray}
    \varphi(\omega;(1,1))
    &=& \frac{-C}{12i} \int_{0}^{\epsilon'} \frac{(1+2u/9)^{-1/2}}{u+\omega} du
    = \frac{-C}{12i} \int_{0}^{\epsilon'} \frac{1-u/9+O(u^{2})}{u+\omega} du
    \nonumber\\
    &\simeq& \frac{-C}{12i} \left[ 
      - \ln|\omega| - \frac{\omega}{9}\ln|\omega| + \cdots \right].
\end{eqnarray}
A similar computation gives
\begin{equation}
    \varphi(\omega;(-1,-1))
    \simeq \frac{-C}{12i} \left[ 
      \ln|\omega| - \frac{\omega}{9}\ln|\omega| + \cdots \right].
\end{equation}
Coming back to the expression (\ref{eq:toy-line-cos}),
the leading order $\ln|\omega|$ is cancelled for the observable
$A_{1}=\cos(\theta_{1}+\theta_{2})$.
However, the second leading order $\omega\ln|\omega|$ survives,
and gives a $t^{-2}$ asymptotic damping.

Similarly, we can find another cancellation for the observable
$A_{4}=\sin(\theta_{1}-\theta_{2})$ on the vertex singularity.
From the above discussions, we build Table~\ref{tab:toy-theory}
which summarizes existing, dominant and cancelled singularities
for the four observables.

\begin{table}
    \centering
    \begin{tabular}{l|cccc}
        \hline
        {\bf Singularity}& {\bf Vertex} & {\bf Tangent} & {\bf Line(leading)} & {\bf Line(second)} \\
        \hline
        {\bf Damping} & $t^{-2}$ & $e^{-it/2}t^{-1.5}$ & $t^{-1}$ & $t^{-2}$ \\
        {\bf Modes} & $(\pm 1,\mp 1)$ & $(\pm 1,\mp 1)$
        & $(\pm 1,\pm 1)$ & $(\pm 1,\pm 1)$ \\
        $(J_{1}^{\ast},J_{2}^{\ast})$ & $(0,0)$ & $(1,0)$ & 
        $J_{2}$ axis & $J_{2}$ axis \\ 
        \hline
        $A_1=\cos(\theta_{1}+\theta_{2})$ & N/A & N/A & C & D \\
        $A_2=\sin(\theta_{1}+\theta_{2})$ & N/A & N/A & D & C \\
        $A_3=\cos(\theta_{1}-\theta_{2})$ & E & D & N/A & N/A \\
        $A_4=\sin(\theta_{1}-\theta_{2})$ & C & D & N/A & N/A \\
        \hline
    \end{tabular}
    \caption{Theoretical prediction of the asymptotic damping
      for the $A_i$ observables in the toy model (\ref{eq:toy-Omega}).
      ``D'' indicates the dominant singularity,
      ``E'' an existing but subdominant singularity,
      and ``C'' a cancelled singularity.
      ``N/A'' tells that the corresponding singularity
      cannot be captured by the observable.
      An oscillation is predicted for the tangent singularity only.}
    \label{tab:toy-theory}
\end{table}

\subsubsection{Numerical computations of the exact solution}

\begin{figure}[h]
    \centering
    \includegraphics[width=7cm]{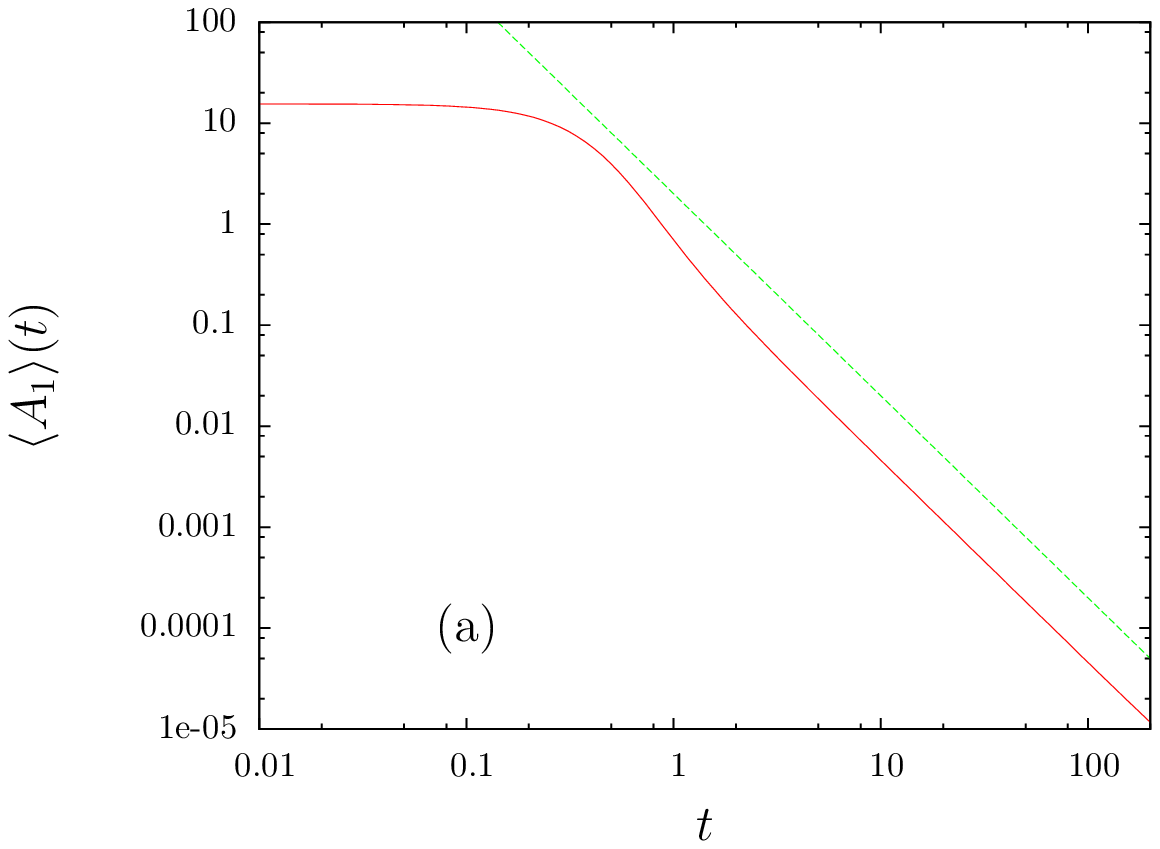}
    \includegraphics[width=7cm]{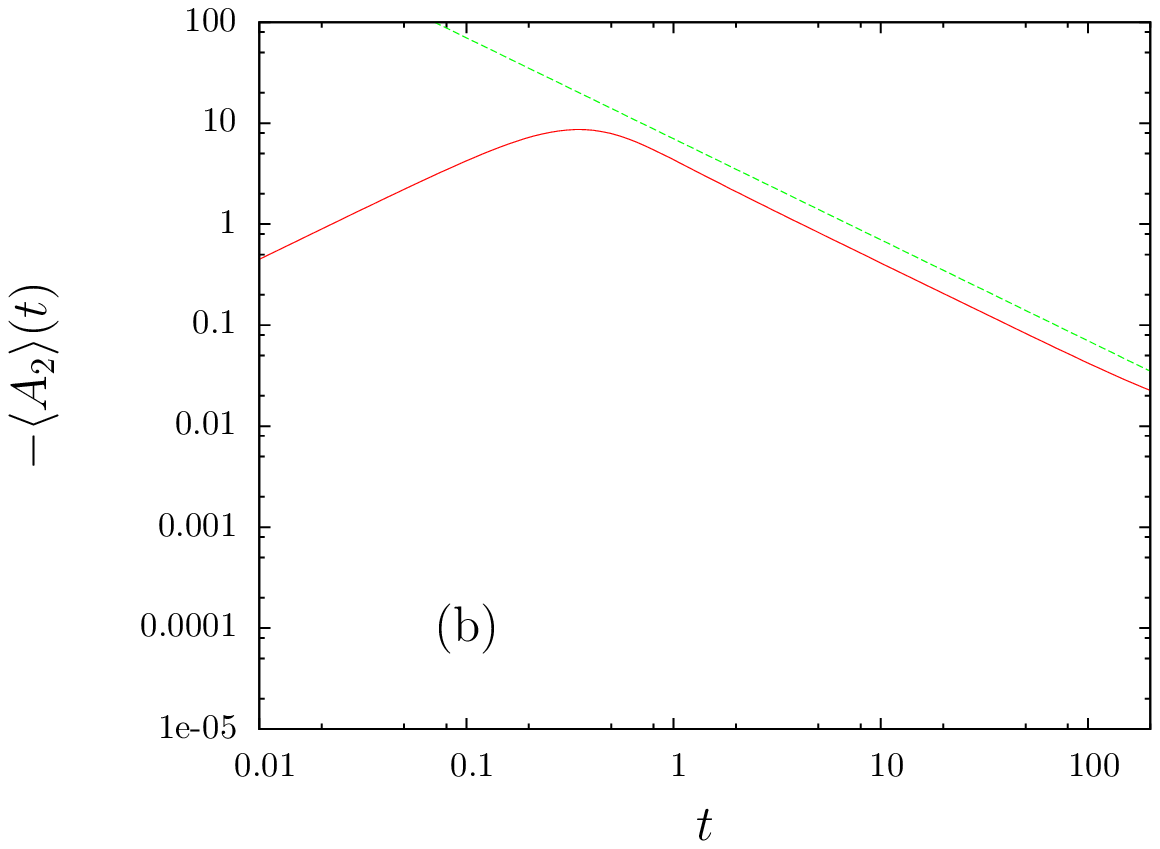}
    \includegraphics[width=7cm]{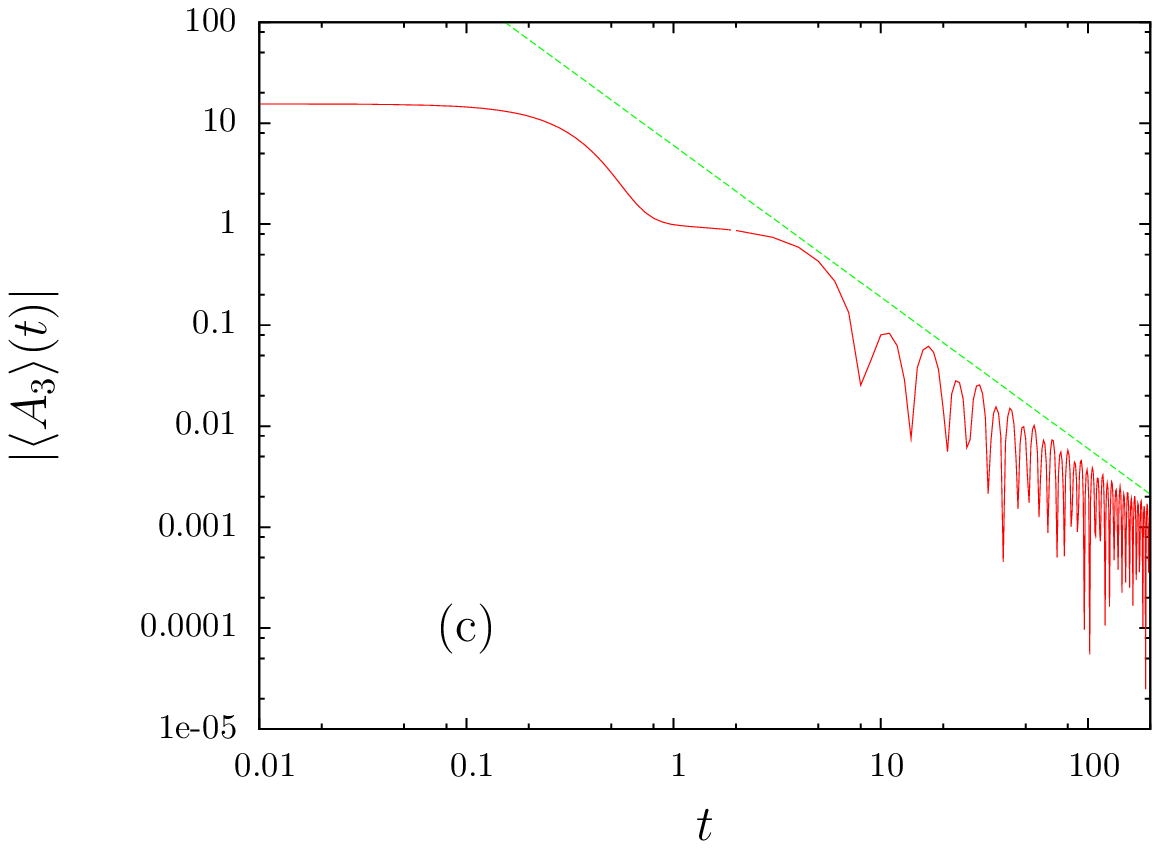}
    \includegraphics[width=7cm]{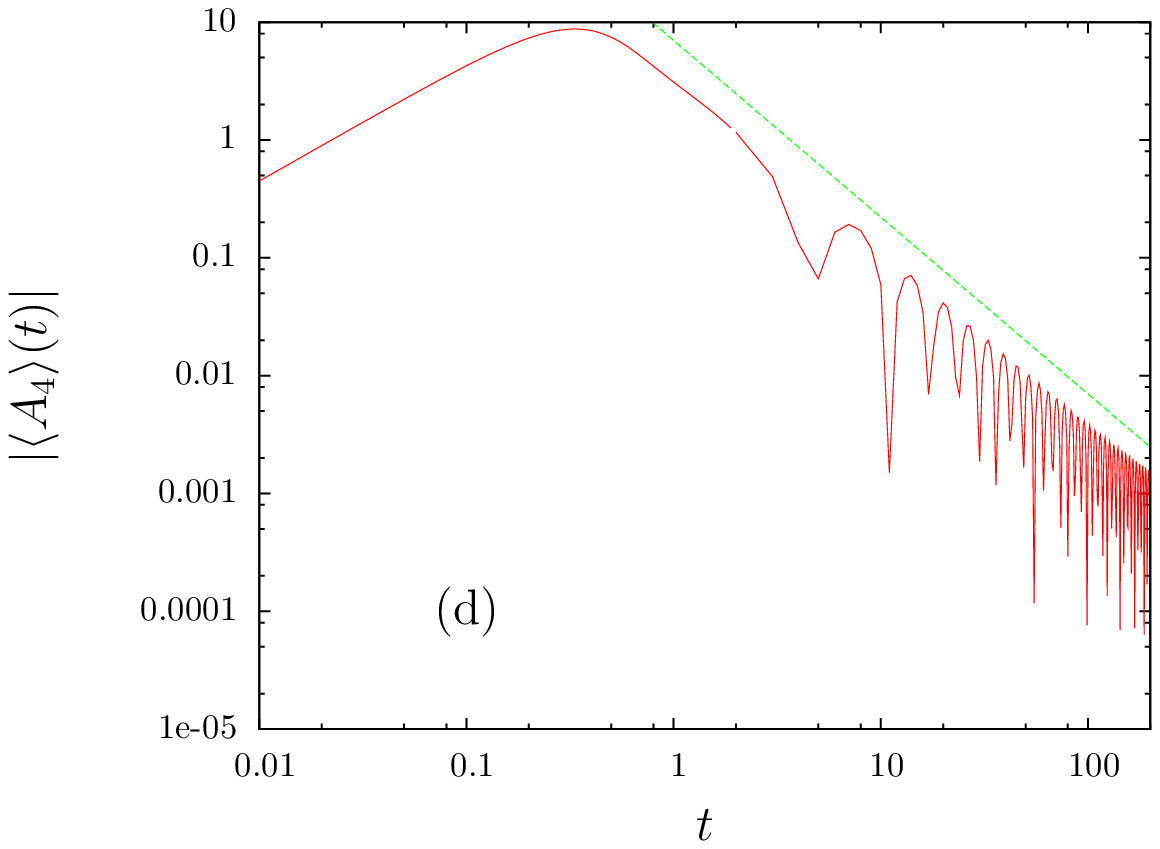}
    \caption{Temporal evolutions of the four observables:
      (a) $\langle A_{1}\rangle(t)$ with $A_{1}=\cos(\theta_{1}+\theta_{2})$.
      (b) $-\langle A_{2}\rangle(t)$ with $A_{2}=\sin(\theta_{1}+\theta_{2})$.
      (c) $|\langle A_{3}\rangle(t)|$ with $A_{3}=\cos(\theta_{1}-\theta_{2})$.
      (d) $|\langle A_{4}\rangle(t)|$ with $A_{4}=\sin(\theta_{1}-\theta_{2})$.
      All panels have the same scales.
      Straight lines show the theoretically predicted
      algebraic dampings, which are respectively 
      $t^{-2},t^{-1},t^{-1.5}$ and $t^{-1.5}$.
      $\langle A_{j}\rangle(t)$ is computed
      by using (\ref{eq:toy-whatwecompute}) with cutoff at
      $J_{1}=J_{2}=10$.}
    \label{fig:toy-result}
\end{figure}

\begin{figure}
    \centering
    \includegraphics[width=7cm]{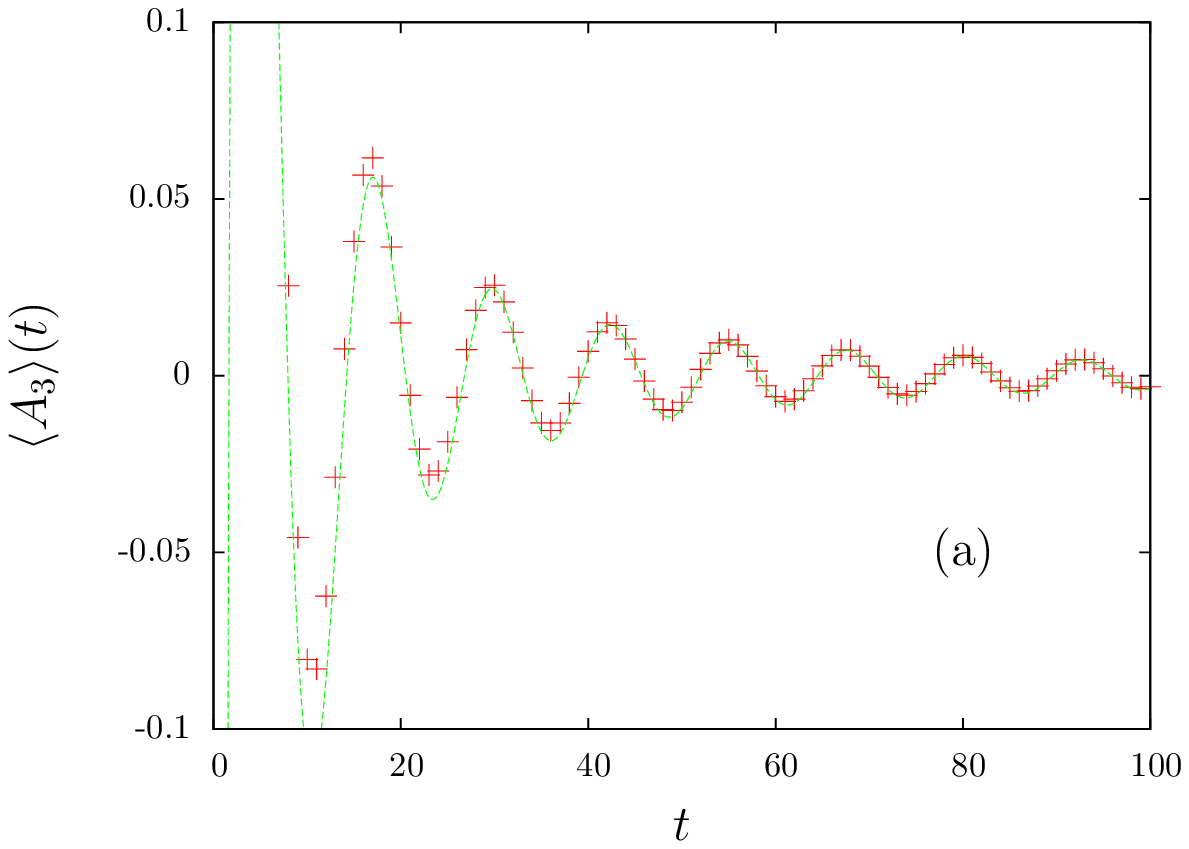}
    \includegraphics[width=7cm]{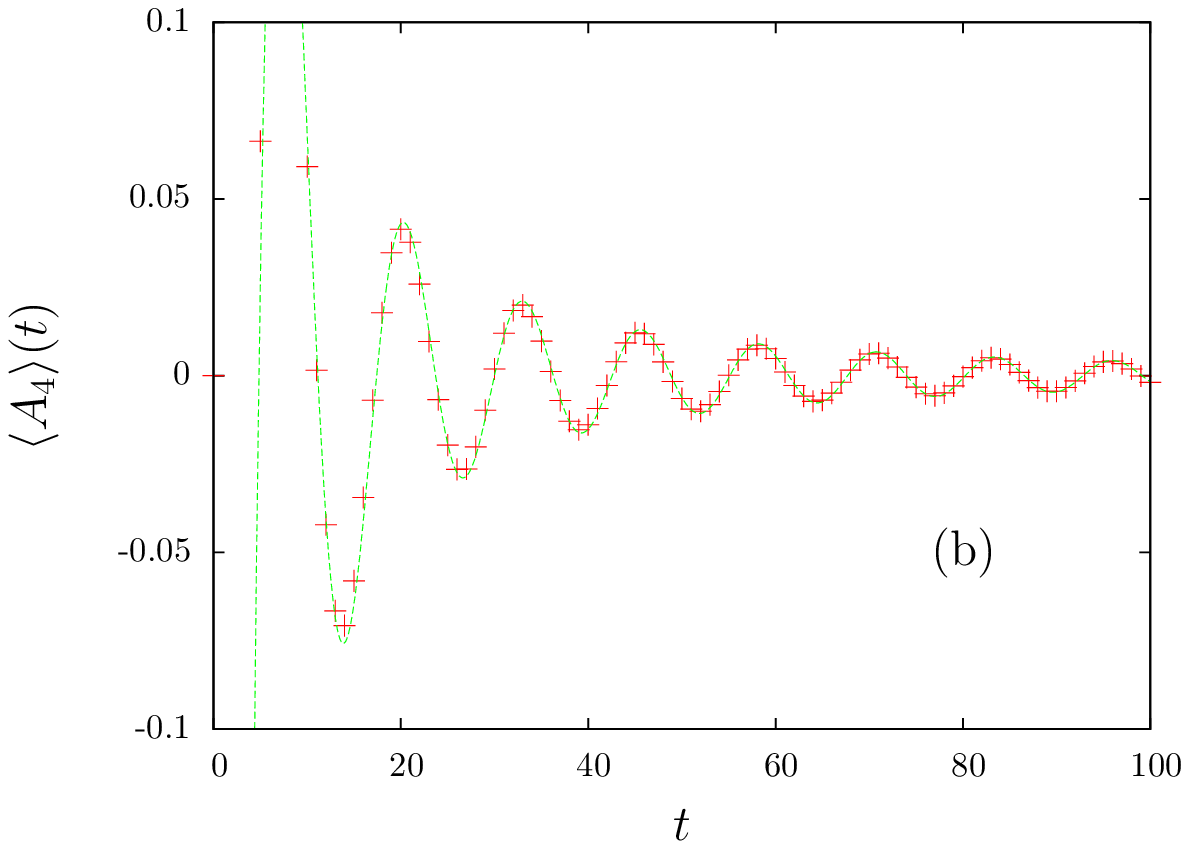}
    \caption{ Temporal evolutions of
      (a) $\langle A_{3} \rangle$ and (b) $\langle A_{4} \rangle$
      in linear scale. 
      Red plus points are numerically obtained,
      and green lines represent $4t^{-1.5}\cos(0.5(t-4.8))$
      and $4t^{-1.5}\cos(0.5(t-8))$ in (a) and (b) respectively.}
    \label{fig:toy-linear}
\end{figure}

On figure~\ref{fig:toy-result}, we compare the theoretical prediction
of Table~\ref{tab:toy-theory}
with the numerically computed exact temporal evolution of the four observables  
(\ref{eq:four-observables}).
We examine both amplitude's damping and frequency of
$\left\langle A_{j} \right\rangle(t)$,
whose exact solution is 
\begin{equation}
    \label{eq:toy-whatwecompute}
    \left\langle A_{j} \right\rangle(t)
    = \pi^{2} \int_{0}^{\infty} \int_{0}^{\infty}
    e^{-(J_{1}^{2}+J_{2}^{2})/2}
    A_{j}(\Omega_{1}(\bJ)t, \Omega_{2}(\bJ)t)
    dJ_{1} dJ_{2}
\end{equation}
by (\ref{eq:exact-expected-value}).
Temporal evolutions of the four observables are reported
in figure~\ref{fig:toy-result},
and the agreements in damping rates are very good
for all observables.
Observables $A_{1}$ and $A_{2}$ have no tangent singularity,
and hence the prediction for the frequencies of
$\left\langle A_{1} \right\rangle(t)$
and $\left\langle A_{2} \right\rangle(t)$ is zero.
These zero frequencies are observed in figures \ref{fig:toy-result}(a) and (b).
On the other hand, the observables $A_{3}$ and $A_{4}$ are
dominated by the tangent singularity at $(J_{1}^{\ast},J_{2}^{\ast})=(1,0)$,
and the prediction for the frequencies is
$|\Omega_{1}(1,0)-\Omega_{2}(1,0)|=0.5$.
Looking at the temporal evolutions in linear scale
in figure~\ref{fig:toy-linear}, we see that the predicted frequency is in
good agreement with the numerics.

\section{Spherically symmetric stellar system}
\label{sec:self-grav}

\subsection{Isochrone model}
We present the isochrone model,
which is a simple example of a spherically symmetric stellar system.
The potential of this model is
\begin{equation}
    \label{eq:isochrone-potential}
    \Phi(r) = - \frac{GM}{b+\sqrt{b^{2}+r^{2}}},
\end{equation}
where $G$ is the gravitational constant,
$M$ is the total mass of the system,
and $b$ is a parameter.
A stationary state satisfying the self-consistent condition
for the potential is known to be~\cite{BinneyTremaine}:
\begin{eqnarray}
    f_{0}
    &=& \left. \frac{1}{\sqrt{2}(2\pi)^{3}(GMb)^{3/2}}
      \frac{\sqrt{\tilde{E}}}{[2(1-\tilde{E})]^{4}}
      \right[
      27 - 66\tilde{E} + 320\tilde{E}^{2}
      - 240\tilde{E}^{3}  \nonumber\\
    && \hspace*{3em}
    \left.
      + 64\tilde{E}^{4} + 3(16\tilde{E}^{2}+28\tilde{E}-9)
      \frac{\sin^{-1}\sqrt{\tilde{E}}}{\sqrt{\tilde{E}(1-\tilde{E})}}
    \right],
    \label{eq:isochrone-stationary}
\end{eqnarray}
where $\tilde{E}=-Eb/GM$ and $E$ is energy.

Let $(r,\vartheta,\psi)$ be the $3$-dimensional polar coordinate,
and $(J_{r},J_{\vartheta},J_{\psi})$ the conjugate angular momenta.
We introduce new actions $(J_{1},J_{2},J_{3})$:
\begin{equation}
    J_{1}=J_{\psi}, \quad J_{2}=J_{\vartheta}+|J_{\psi}|, \quad
    J_{3}=J_{r}.
\end{equation}
We denote the conjugate angle variables by $(\theta_{1},\theta_{2},\theta_{3})$.
The Hamiltonian of this model is then 
\begin{equation}
    \label{eq:isochrone-hamiltonian}
    H(\bJ) = - \frac{(GM)^{2}}{2[J_{3}+\frac{1}{2}(J_{2}+\sqrt{J_{2}^{2}+4GMb})]^{2}}. 
\end{equation}
The Hamiltonian is defined on the domain
$(J_{1},J_{2},J_{3})\in [-J_{2},J_{2}]\times [0,\infty)\times [0,\infty)$.
We remark that the Hamiltonian depends on $J_{2}$ and $J_{3}$,
but not on $J_{1}$.
We hence omit the modes corresponding to $J_{1}$,
and write $\bm=(m_{2},m_{3})$ for the mode vector
and $\bOmega=(\Omega_{2},\Omega_{3})$ for the frequency vector.
Thus, the isochrone model fits in the $2$-dimensional analysis
developed in this paper.
Note that for consistency with standard notations,
the actions $J_{2}$ and $J_{3}$ in the isochrone model
correspond to $J_{1}$ and $J_{2}$ in the general theory
of section~\ref{sec:singularities-FG}.
Frequencies are given by
\begin{eqnarray}
    \label{eq:isochrone-Omega3}
    \Omega_{3}(\bJ)
    &=& \frac{\partial H}{\partial J_{3}}(\bJ)
    = \frac{(GM)^{2}}{\left[J_{3}+\frac{1}{2}(J_{2}+\sqrt{J_{2}^{2}+4GMb}) \right]^{3}}, \\
    \label{eq:isochrone-Omega2}
    \Omega_{2}(\bJ)
    &=& \frac{\partial H}{\partial J_{2}}(\bJ)
    = \frac{1}{2} \left( 1 + \frac{J_{2}}{\sqrt{J_{2}^{2}+4GMb}} \right)
    \Omega_{3}(\bJ).
\end{eqnarray}

\subsection{Advection equation corresponding to isochrone model}

Using the frequencies $\Omega_{2}$ (\ref{eq:isochrone-Omega2})
and $\Omega_{3}$ (\ref{eq:isochrone-Omega3}),
we consider the advection equation
\begin{equation}
    \label{eq:isochrone-advection}
    \frac{\partial f_{1}}{\partial t}
    + \Omega_{2}(\bJ)
    \frac{\partial f_{1}}{\partial \theta_{2}}
    + \Omega_{3}(\bJ)
    \frac{\partial f_{1}}{\partial \theta_{3}} = 0.
\end{equation}
This advection equation is obtained by omitting
the potential perturbation in the linearized Vlasov equation.
We take as initial state:
\begin{equation}
    f(\bt,\bJ,t=0) = f_{0}(\bJ) + f_{1}(\bt,\bJ,t=0),
\end{equation}
where $f_{0}(\bJ)$ is the stationary state (\ref{eq:isochrone-stationary})
and the perturbation $f_{1}$ is set as
\begin{equation}
    \label{eq:isochrone-perturbation}
    f_{1}(\bt,\bJ,t=0) = a f_{0}(\bJ) \cos(n_{2}\theta_{2})\cos(n_{3}\theta_{3}),
    \quad |a|\ll 1.
\end{equation}
Note that this perturbation is also independent of $\theta_{1}$ and $J_{1}$.
The Fourier transform $ig(\bm,\bJ)$ of $f_{1}(\bt,\bJ,t=0)$ is then
\begin{equation}
    i g(\bm,\bJ) = \left\{
      \begin{array}{lll}
          a f_{0}(\bJ)/4 & (m_{2},m_{3})=(\pm n_{2},\pm n_{3}) \\
          0 & {\mbox otherwise.}
      \end{array}
    \right.
\end{equation}

Let us consider the temporal evolution of the expected value
of an observable $A(\theta_{2},\theta_{3})$.
As discussed in section \ref{sec:advection}, the Laplace transform
of the expected value is given by (\ref{eq:fourier-laplace-A}),
but the integrals should be performed over
$\theta_{1},\theta_{2},\theta_{3},J_{1},J_{2}$ and $J_{3}$.
Remembering that $A$, $f_{1}$ and $\bOmega$ do not depend on
$\theta_{1}$ and $J_{1}$,
the expected value may be written as a linear combination of functions of the form
\begin{equation}
    \label{eq:isochrone-varphi}
    \varphi(\omega;\bm)
    = \int_{0}^{\infty} dJ_{3} \int_{0}^{\infty} dJ_{2}
    ~\frac{1}{\bm\cdot\bOmega(\bJ)-\omega}
    \int_{-J_2}^{J_2}h(\bm,\bJ)dJ_1.
\end{equation}
Note that if we were studying the true linearized Vlasov equation, the
function $h$ in the above equation would be replaced by
$\bm\cdot\nabla_{\bJ}f_{0}(\bJ)\bar{c}_{l}(\bm,\bJ)c_{k}(\bm,\bJ)$ or
$g(\bm,\bJ)\bar{c}_{l}(\bm,\bJ)$, according to the definitions of
$F_{lk}(\omega)$ and $G_{l}(\omega)$ in equations (\ref{eq:F}) and
(\ref{eq:G}) respectively. Since $\int_{-J_2}^{J_2}hdJ_1$ vanishes
for $J_2=0$, a $J_2$ factor can be factorized, and we can write
$\int_{-J_2}^{J_2}h(\bm,\bJ)dJ_1=J_2\tilde{h}(\bm,\bJ)$.  Thus, the
functions (\ref{eq:isochrone-varphi}) fit in the framework of the
abstract setting (\ref{eq:varphi}), and we apply the theory developed
in sections~\ref{sec:singularities-FG} and \ref{sec:asymptotic}, with
\begin{equation}
    \label{eq:isochrone-numu}
    \nu(\bJ) = J_{2}\tilde{h}(\bm,\bJ), \quad
    \denom(\bJ) = \bm\cdot\bOmega(\bJ).
\end{equation}
The $J_2$ factor in $\nu(\bJ)$
will be of importance to determine the singularity associated 
with vertex, line and infinite singularities.

\subsection{Singularities in isochrone model and theoretical prediction}
\label{sec:isochrone-singularity}

We can now list the singularities in the isochrone model:
\begin{itemize}
      \item Vertex singularity at $(J_{2}^{\ast},J_{3}^{\ast})=(0,0)$, 
    except for the modes $\bm$ satisfying
    $m_{2}+2m_{3}=0$ or $m_{2}+3m_{3}=0$.
    The former case results in a line singularity,
    and the latter case gives a special vertex singularity
    with $\partial(\bm\cdot\bOmega)/\partial J_{2}=0$ at the origin.
      \item Tangent singularity at $(J_{2}^{\ast},0)$
    for the modes $\bm$ satisfying $-1<m_{3}/m_{2}<-1/3$.
    For such a mode, the singular point $J_{2}^{\ast}$ is given
    by the solution to the equation
    \begin{equation}
        \left. \frac{\partial}{\partial J_{2}} \right|_{J_{3}=0}
        \bm\cdot\bOmega = 0,
    \end{equation}
    which reads
    \begin{equation}
        \label{eq:dmOmegadJ2=0}
        -\frac{1}{2}\left( 1+ \frac{J_{2}}{\sqrt{J_{2}^{2}+4GMb}} \right)
          + \frac{2}{3} \frac{GMb}{J_{2}^{2}+4GMb}
        = \frac{m_{3}}{m_{2}} .
    \end{equation}
    We remark that the left-hand-side of (\ref{eq:dmOmegadJ2=0})
    is a decreasing function of $J_{2}$,
    and its range is $(-1,-1/3)$ for $J_{2}\in (0,\infty)$.
    Thus $m_{3}/m_{2}$ must be in the interval $(-1,-1/3)$.
    The origin, namely $J_{2}^{\ast}=0$, results in the special vertex singularity
    for the modes satisfying $m_{2}+3m_{3}=0$.
      \item Line singularity on $J_{3}$ axis for the modes $\bm$
    satisfying $m_{2}=-2m_{3}$.
      \item A singularity at infinity appears for any mode $\bm$.
    Estimations of $\nu(\bJ)$ and $\denom(\bJ)$ are
    \begin{equation}
        \nu(\bJ) \sim (J_{2}+J_{3})^{-4}, \quad
        \denom(\bJ) \sim (J_{2}+J_{3})^{-3},
    \end{equation}
    since $\Omega_{2}\sim\Omega_{3}\sim (J_{2}+J_{3})^{-3}$,
    $f_{0}\sim \tilde{E}^{5/2}$, $\tilde{E}\sim (J_{2}+J_{3})^{-2}$,
    and taking into account the $J_{2}$ factor for $\nu$ appearing in
    (\ref{eq:isochrone-numu}).
\end{itemize}
No critical singularity appears in the isochrone model.
These singularities are arranged in Table~\ref{tab:singularity-isochrone}
with asymptotic dampings and frequencies.

\begin{table}
    \centering
    \begin{tabular}{lccc}
        \hline
        {\bf Singularity} & {\bf Damping} & {\bf Frequency} & {\bf Condition for mode} \\
        \hline
        Vertex & $t^{-3}$ & $\left(\frac{m_{2}}{2}+m_{3}\right)\frac{\sqrt{GM}}{2^{4}b^{3/2}}$ & None \\
        Tangent & $t^{-1.5}$ & $m_{2}\Omega_{2}(J_{2}^{\ast},0)+m_{3}\Omega_{3}(J_{2}^{\ast},0)$ & $-1<m_{3}/m_{2}<-1/3$ \\
        Line & $t^{-2}$ & $0$ & $m_{2}=-2m_{3}$ \\
        Infinity & $t^{-2/3}$ & $0$ & None \\
        \hline
    \end{tabular}
    \caption{Dampings and frequencies associated with
      singularities in the isochrone model.
      This table shows the leading singularity for each type.
      $(m_{2},m_{3})$ represents a mode.
      $J_{2}^{\ast}$ is the solution to (\ref{eq:dmOmegadJ2=0}).
      Modes satisfying $m_{2}+2m_{3}=0$ or $m_{2}+3m_{3}=0$
      do not yield a generic vertex singularity, but a line singularity
      and a special vertex singularity respectively.
      See text for the leading line singularity.}
    \label{tab:singularity-isochrone}
\end{table}

The leading dampings are $t^{-3}$, $t^{-1.5}$, $t^{-2}$ and $t^{-2/3}$
for the vertex, tangent, line and
infinity singularities respectively.
We remark that the leading order of the function $\nu$
is $\nu\sim J_{2}$ around $J_{2}=0$
due to the $J_{2}$ factor of $\nu$ in (\ref{eq:isochrone-numu}),
thus the leading damping rates
for the vertex and line singularities respectively
are not $t^{-2}$ and $t^{-1}$, but $t^{-3}$ and $t^{-2}$ .
The dominant, i.e. slowest, decay is $t^{-2/3}$ with zero frequency.

In the next subsection we will observe the temporal evolution
of the expected value of the observable
$A(\bt)=\sin(n_{2}\theta_{2}+n_{3}\theta_{2})$
with respect to the exact solution
\begin{equation}
    f_{1}(\bt,\bJ,t) = f_{1}(\bt-\bOmega(\bJ)t, \bJ, t=0).
\end{equation}
The mode $(n_{2},n_{3})$ corresponds to
the mode of perturbation (\ref{eq:isochrone-perturbation}).
The expected value is:
\begin{equation}
    \left\langle A \right\rangle_{1}(t)=\int A(\bt)f_{1}(\bt,\bJ,t)d\bt d\bJ.
\end{equation}
For our interests, the prefactor of $\left\langle A \right\rangle_{1}(t)$
is not crucial; therefore we redefine $\left\langle A \right\rangle_{1}(t)$ as
\begin{equation}
    \label{eq:isochrone-whatwecompute}
    \left\langle A \right\rangle_{1}(t)
    = \int_{0}^{\infty} \int_{0}^{\infty} J_{2} f_{0}(\bJ) \sin ( (n_{2}\Omega_{2}(\bJ)+n_{3}\Omega_{3}(\bJ)) t) dJ_{2} dJ_{3}
\end{equation}
in the following.
For this sine observable,
the leading vertex, tangent and infinity singularities survive,
but the leading line singularity is expected to be cancelled
looking at Table~\ref{tab:singularity-alternate},
since the corresponding exponents are $a_{1}=1$ and $a_{2}=0$.
From the line singularity, therefore,
the second leading damping $t^{-3}$ should arise
instead of the leading damping $t^{-2}$.
However, since the line singularity is never dominant,
this cancellation does not affect the following discussions.

\subsection{Numerical check}

\begin{figure}
    \centering
    \includegraphics[width=10cm]{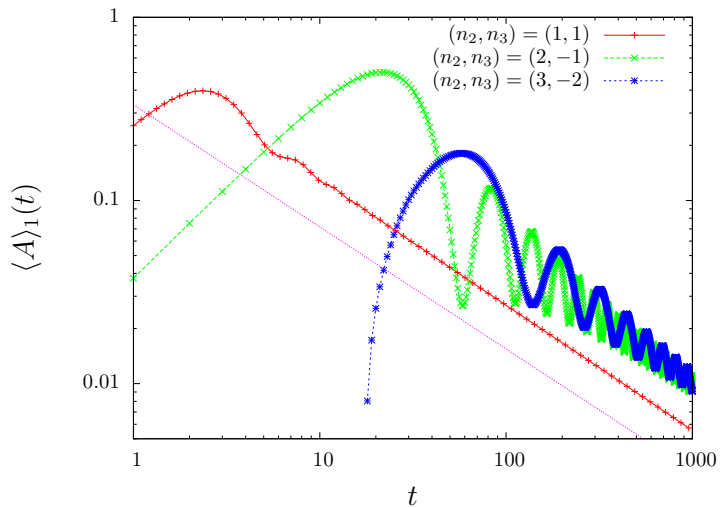}
    \caption{Damping in the isochrone model
      for the observable $A=\sin(n_{2}\theta_{2}+n_{3}\theta_{3})$.
      $(n_{2},n_{3})=(1,1)$ (red plus),
      $(2,-1)$ (green cross) and $(3,-2)$ (blue star).
      The purple guide line is proportional to $t^{-2/3}$.
      $\langle A \rangle(t)$ is computed by using
      (\ref{eq:isochrone-whatwecompute}) with cutoff at $J_{2}=J_{3}=20$.
    \label{fig:isochrone-damping}}
\end{figure}

We choose three modes: $(n_{2},n_{3})=(1,1),(2,-1)$ and $(3,-2)$.
The mode $(1,1)$ has vertex and infinity singularities,
$(2,-1)$ has all four singularities,
and $(3,-2)$ has all four except the line singularity.
To perform the numerical integration of (\ref{eq:isochrone-whatwecompute}),
we set the parameters as $G=M=b=1$.
The exact solutions (\ref{eq:isochrone-whatwecompute})
are shown in figure~\ref{fig:isochrone-damping}.

The mode $(1,1)$ shows $t^{-2/3}$ damping without oscillation
as the theory predicted.
This mode also has a vertex singularity, which gives a non-zero frequency,
but no oscillation is observed in the asymptotic time region:
the vertex singularity contribution is expected to damp as $ t^{-3}$, 
which may be too fast to be visible at large times.

\begin{figure}
    \centering
    \includegraphics[width=10cm]{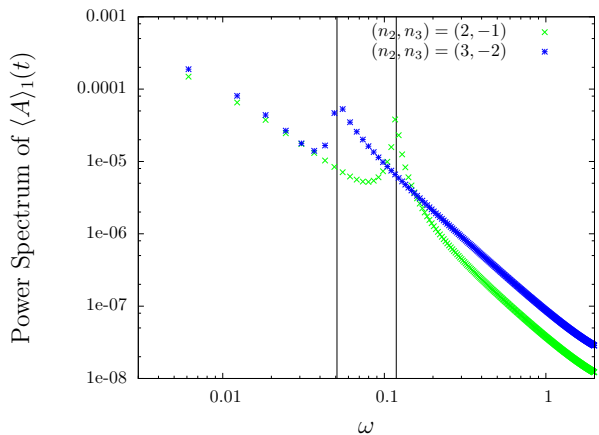}
    \caption{ 
      Power spectra of $\langle A \rangle_{1}(t)$
      for the modes $(2,-1)$ (green cross points) and $(3,-2)$
      (blue star points).
      The two vertical lines indicate $\omega=0.1185$ and $0.0509$,
      which are predicted theoretically 
      for the modes $(2,-1)$ and $(3,-2)$ respectively as
      $|m_{2}\Omega_{2}(J_{2}^{\ast},0)+m_{3}\Omega_{3}(J_{2}^{\ast},0)|$
      with $J_{2}^{\ast}$ the solution to (\ref{eq:dmOmegadJ2=0}).
      The power spectra are computed from time series with time slice $1$
      in the time interval $[77,1100]$ to avoid an early transient time region.}
    \label{fig:isochrone-powerspectra}
\end{figure}

The modes $(2,-1)$ and $(3,-2)$ asymptotically damp as $t^{-2/3}$
as predicted, but they have oscillations.
At variance with the mode $(1,1)$, both these modes have 
a tangent singularity, which gives a rather slow damping $t^{-1.5}$
with non-zero frequency.
To confirm that there is a contribution from the tangent singularity,
we show the power spectra of $\left\langle A \right\rangle_{1}(t)$
in figure~\ref{fig:isochrone-powerspectra}.
The peaks of the power spectra are in good agreement with
the theoretically predicted frequency
$|m_{2}\Omega_{2}(J_{2}^{\ast},0)+m_{3}\Omega_{3}(J_{2}^{\ast},0)|$,
where $J_{2}^{\ast}$ is the solution to (\ref{eq:dmOmegadJ2=0}).

As mentioned above, to study the true Vlasov equation,
we should also estimate the $c_{k}(\bm,\bJ)$ functions,
which depend on the biorthogonal functions $u_{k}(\bq)$.

\section{Conclusion}

When a stable inhomogeneous stationary state of a Vlasov equation is
perturbed, the asymptotic damping of the perturbation is algebraic, no
matter how regular the stationary state and perturbation are. The
damping rate and frequency are controlled by the singularities on the
real axis of the Fourier-Laplace transform of the perturbation. In
this paper, in systems with two spatial dimensions,
we have classified these singularities 
into vertex, tangent, critical, line and infinity singularities,
and have given damping rate and frequency in each case.
This classification is also valid for
three dimensional spherically symmetric stationary states,
since they reduce to the two-dimensional case thanks to symmetry.
The resulting picture is much richer than in one spatial
dimension~\cite{BOY11}.  We have illustrated the theory on a toy
model, and on an advection equation associated with the isochrone
model, a simple model for self-gravitating systems: 
tests in these models show that
the singularity analysis captures very well the observed damping.

The main goal of this theory is a better understanding of the
asymptotic relaxation of self gravitating systems.  To achieve this
goal, a detailed study of the classical expansion of a perturbation on
a biorthogonal basis should be coupled to the singularity analysis we
have presented here.

\ack
The authors thank Magali Ribot for suggesting the power series expansions
(\ref{eq:expansion-u-x}).
This work is partially supported by the ANR-09-JCJC-009401 INTERLOP project.
Y.Y.Y. acknowledges the support of a Grant-in-Aid for Scientific Research (C) 23560069.

\appendix
\section{Analytic continuation}
\label{sec:continuation-details}

The starting point is the following formula for $\varphi(z)$, valid for 
${\mbox Im}(z)>0$:
\[
\varphi(z)=\int_{D\subset\mathbb{R}^2}\frac{\nu(J_1,J_2)}{\mu(J_1,J_2)-z}~.
\]
We want to understand the regularity of $\varphi(z)$ close to 
$z=x_0 \in\mathbb{R}$.

\begin{figure}
\centering{\includegraphics[width=8cm]{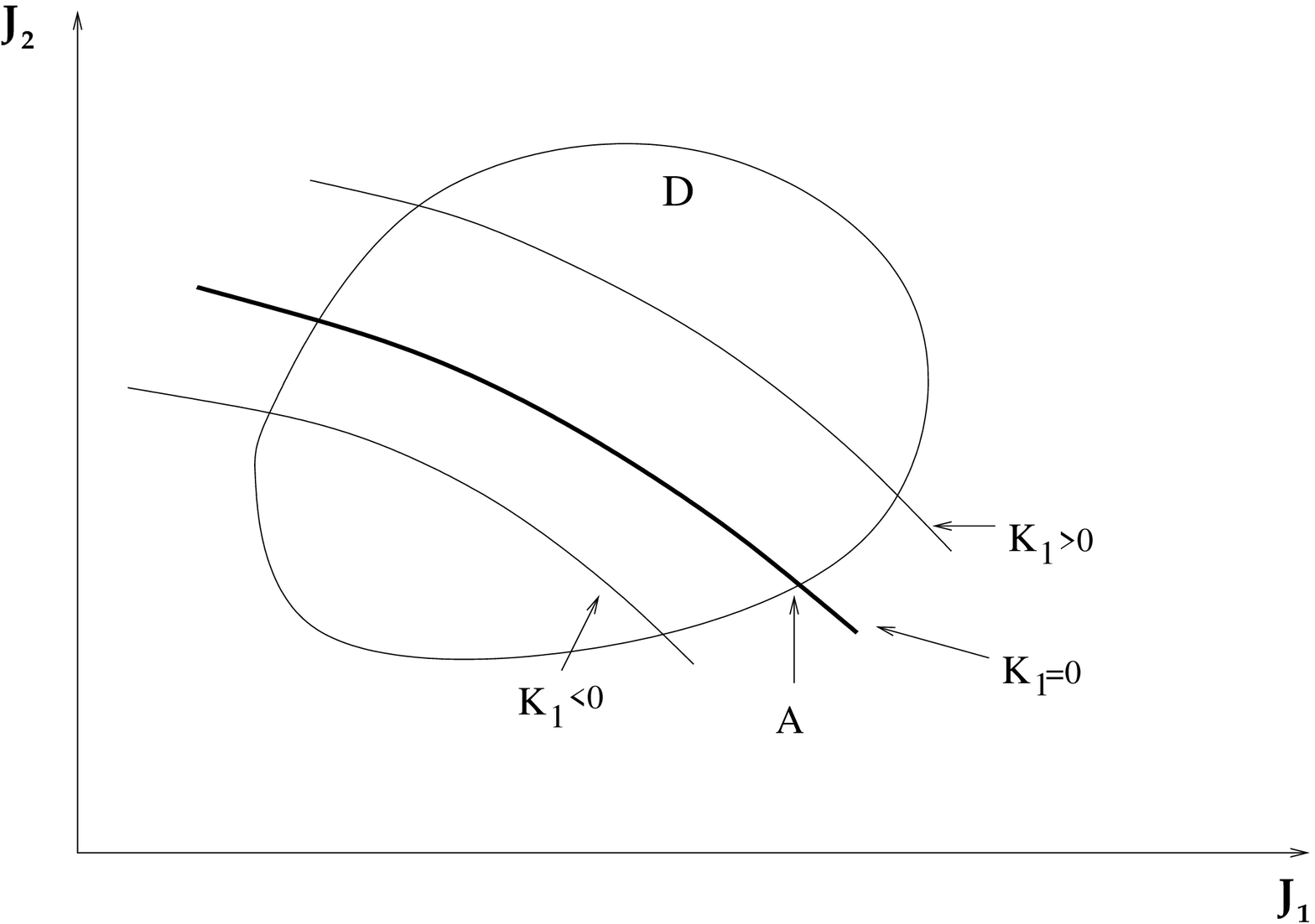}}
\caption{\label{fig2} The boundary of $D$ is regular close to point 
$A$, its intersection with the curve $\mu(J_1,J_2)=x_0$, that is $K_1=0$}.
\end{figure}
Assuming that $x_0$ is not a critical value for the function $\mu$ 
(condition i), the 
equation $\mu(J_1,J_2)=x_0$ defines one or several curves on the
$(J_1,J_2)$ plane. We treat here the case with one curve, the argument
carries over easily to the case of several curves.

It is then possible to use a new set of variables $K_1,K_2$ fulfilling the 
two conditions:\\
a) $K_1=\mu(J_1,J_2)-x_0$ \\
b) $K_2$ is such that the Jacobian $dK_1dK_2/dJ_1dJ_2$ never 
vanishes. We do not need to further specify $K_2$.

Then, with a suitable definition of $\tilde{\nu}$:
\[
\varphi(z)=\int_{\tilde{D}\subset\mathbb{R}^2}\frac{\tilde{\nu}(K_1,K_2)}
{K_1-(z-x_0)}dK_1dK_2~.
\]
Integrating over $K_2$ first:
\[
\varphi(z)=\int_{K_1^{(m)}}^{K_1^{(M)}}\frac{dK_1}{K_1-(z-x_0)} 
\int_{K_2^{(m)}(K_1)}^{K_2^{(M)}(K_1)} \tilde{\nu}(K_1,K_2)dK_2~.
\]
If the boundary of the domain $D$ is regular around its intersection with 
the curve $K_1=0$ (condition ii) 
(see figure~\ref{fig2}, point~$A$), then $K_2^{(M)}(K_1)$ and $K_2^{(m)}(K_1)$ 
are regular functions of $K_1$, and we are left with the single integral
\[
\varphi(z)=\int_{K_1^{(m)}}^{K_1^{(M)}}\frac{h(K_1)}{K_1-(z-x_0)} 
dK_{1}
\]
where $h$ is a regular function. The usual Landau continuation 
argument then ensures that $\varphi(z)$ is regular at $z=x_0$, 
unless $K_1^{(m)}=0$ or $K_1^{(M)}=0$ (condition iii).

Thus, under conditions i), ii) and iii), $\varphi(z)$ is not singular
at $z=x_0$. Breaking one of these conditions yields a
critical (section~\ref{sec:critical}), vertex (section~\ref{sec:vertex}) and
tangent (section~\ref{sec:tangent}) singularity respectively.

\section{Vertex singularity}
\label{sec:vertex-singularity-details}
\begin{figure}
    \centering
    \includegraphics[width=15cm]{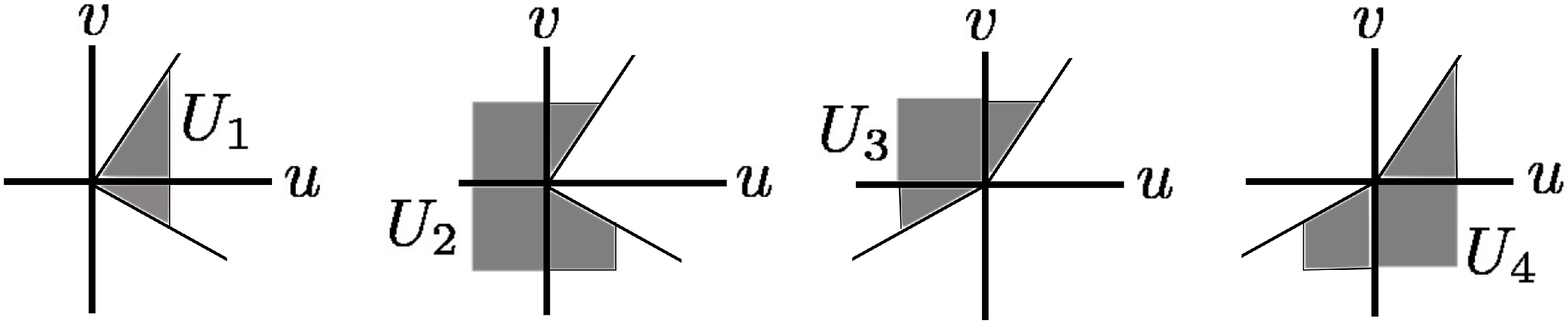}
    \caption{Four types of domain $U$ for the vertex singularity.}
    \label{fig:four-domain}
\end{figure}

Let us compute the leading singularity of
\begin{equation}
    \label{eq:appendix-vertex-singularity}
    \varphi(z) = \int_{U} \frac{u^{k}v^{l}}{u-(z-x_{0})} dudv. 
\end{equation}
The origin $(u,v)=(0,0)$ is on the boundary of the domain $U$,
and is a singular point.
We assume that the
boundary of the domain $U$ is locally approximated
by two lines around the origin,
and that the line $u=0$ does not coincide
with one of such boundary lines.
We can classify the shapes of the domain in three cases:
(i) the two lines are in the $u>0$ half plane;
(ii) the two lines are in the $u<0$ half plane;
(iii) one line is in the $u>0$ half plane,
and the other in the $u<0$ half plane.
Case (ii) is made identical to case (i) by changing the sign of
$u,z$ and $C$ if necessary.
Thus we consider case (i) and case (iii).
Remembering that the line $u=0$ does not coincide with the boundary
of $U$, we can further classify $U$ 
into the following four types:
\begin{equation}
    U_{1} = \{ (u,v) ~|~ 0<u<\epsilon,~ \beta u<v<\alpha u\}.
\end{equation}
\begin{eqnarray}
    U_{2}
    &=&
    \{ (u,v) ~|~ 0<u<\epsilon,~ \alpha u<v<\delta {\rm ~or~ }\delta<v<\beta u\} \nonumber\\
    && \cup \{ (u,v) ~|~ -\epsilon<u<0,~ -\delta<v<\delta\}.
\end{eqnarray}
\begin{eqnarray}
    U_{3}
    &=& \{ (u,v) ~|~ 0<u<\epsilon,~ \alpha u<v<\delta\}\nonumber\\
    && \cup \{ (u,v) ~|~ -\epsilon<u<0,~ \beta u<v<\delta\}.
\end{eqnarray}
\begin{eqnarray}
    U_{4}
    &=& \{ (u,v) ~|~ 0<u<\epsilon,~ \delta <v<\alpha u\}\nonumber\\
    && \cup \{ (u,v) ~|~ -\epsilon<u<0,~ \delta <v<\beta u\}.
\end{eqnarray}
The domains $U_{1}$ and $U_{2}$ correspond to case (i),
and $U_{3}$ and $U_{4}$ to case (iii).
$\delta$, $\epsilon$, $\alpha$ and $\beta$ are constants.
See figure~\ref{fig:four-domain} for schematic pictures of the domains.
We first note that the integral over the domain
\begin{equation}
    U_{\rm regular} = \{ (u,v)~|~ -\epsilon<u<\epsilon, ~ \alpha u<v<\delta \}
\end{equation}
does not give any singularity for the function
(\ref{eq:appendix-vertex-singularity}).
This fact is immediately obtained by performing the integral over $v$:
\begin{equation}
    \int_{-\epsilon}^{\epsilon} \frac{u^{k}du}{u-(z-x_{0})} \int_{\alpha u}^{\delta}
    v^{l} dv
    = \int_{-\epsilon}^{\epsilon} \frac{u^{k}}{u-(z-x_{0})}
    ~\frac{\delta^{l+1}-(\alpha u)^{l+1}}{l+1} ~ du
\end{equation}
and using \ref{sec:main_computation}.
Adding or subtracting such a regular domain,
domains $U_{2},U_{3}$ and $U_{4}$ can be reduced to $U_{1}$.
Thus, what we have to consider are integrals of the form:
\begin{equation}
    \varphi(z) = \int_{0}^{\epsilon} \frac{u^{k}du}{u-(z-x_{0})}
    \int_{\beta u}^{\alpha u} v^{l} dv
    = \frac{\alpha^{l+1}-\beta^{l+1}}{l+1}
    \int_{0}^{\epsilon} \frac{u^{k+l+1}}{u-(z-x_{0})} du
\end{equation}
Using \ref{sec:main_computation} again,
the leading singularity of $\phi(x)=\lim_{y\to 0^+} \varphi(x+iy)$ 
at $x_{0}$ is 
\begin{equation}
    \phi^{\rm sing}_{x_{0}}(x)
    = c_{1} (x-x_{0})^{k+l+1}\ln |x-x_{0}| + c_{2} (x-x_{0})^{k+l+1} H(x-x_{0})
\end{equation}
except in the non-generic situation $\beta=-\alpha$, $l$ odd.

\section{Tangent singularity}
\label{sec:tangent-singularity-details}

Changing variables to $u=J_{1}^{2}+J_{2}, v=J_{1}$,
the leading singular part of the function $\varphi(z)$ is:
\begin{eqnarray}
    \varphi(z)
    &=& C \int_{0}^{u_{0}} \frac{du}{u-(z-x_{0})}
    \int_{-\sqrt{u}}^{\sqrt{u}} v^{a_{1}}(u-v^{2})^{a_{2}} dv\\
    &=& \sum_{k} c_{k} \int_{0}^{u_{0}} \frac{u^{a_{2}-k} du}{u-(z-x_{0})}
    \int_{-\sqrt{u}}^{\sqrt{u}} v^{a_{1}+2k} dv .
\end{eqnarray}
The integral over $v$ is computed as
\begin{displaymath}
    \int_{-\sqrt{u}}^{\sqrt{u}} v^{a_{1}+2k} dv
    = \frac{1}{1+a_{1}+2k} ( 1 - (-1)^{1+a_{1}+2k})
      u^{(1+a_{1}+2k)/2}.
\end{displaymath}
Thus we have
\begin{displaymath}
    \varphi(z)
    = \sum_{k} c_{k} \frac{1 - (-1)^{1+a_{1}+2k}}{1+a_{1}+2k}
    \int_{0}^{u_{0}} \frac{u^{1/2+a_{1}/2+a_{2}}}{u-(z-x_{0})}du.
\end{displaymath}
The prefactor $1-(-1)^{1+a_{1}+2k}=1-(-1)^{1+a_{1}}$ vanishes
for any $k$ when $a_{1}$ is odd,
thus we may assume that $a_{1}$ is even. We are left with integrals
as in (\ref{eq:singularity-computation-function1}),
with $1/2+a_1/2+a_2 \notin \mathbb{Z_+}$. Thus the leading singularity is
\begin{equation}
    \phi^{\rm sing}_{x_{0}}(x)
    = C_{1}(x-x_{0})^{1/2+a_{1}/2+a_{2}} H(x-x_{0})
    + C_{2}(x_{0}-x)^{1/2+a_{1}/2+a_{2}} H(x_{0}-x).
\end{equation}

\section{Critical singularity}
\label{sec:critical_appendix}
Calling $\lambda_{1}$ and $\lambda_{2}$ 
the two eigenvalues of the Hessian of $\mu$ at the critical point,
we expand $\denom(\bJ)$ after an appropriate
change of basis
\begin{equation}
    \denom(\bJ)
    \simeq x_{0} +\lambda_{1} (J_{1}-J_{1}^{\ast})^{2}
    + \lambda_{2}(J_{2}-J_{2}^{\ast})^{2}.
\end{equation}
We have to consider two cases: i) the critical point is a local extremum;
ii) the critical point is a saddle.

\paragraph{Case i), $\lambda_1\lambda_2>0$}: 
Without loss of generality, we assume $\lambda_1>0~,\lambda_2>0$.
Using the leading order approximation~(\ref{eq:g-leading})
for $\nume$ and performing the usual shifting and scaling of variables,
we obtain the leading singular part of $\varphi(z)$ as
\begin{eqnarray}
    \varphi(z)
    &=& C \int_{U}
    \frac{J_{1}^{a_{1}}J_{2}^{a_{2}}}{J_{1}^{2}+J_{2}^{2}-(z-x_{0})}dJ_{1}dJ_{2}  \\
    &=& C \int_{0}^{2\pi} \cos^{a_{1}} \alpha \sin^{a_{2}}\alpha
    d\alpha \int_{0}^{\varepsilon} \frac{r^{1+a_{1}+a_{2}}}{r^{2}-(z-x_{0})}dr
\end{eqnarray}
where $(r,\alpha)$ are the polar coordinates on the $(J_{1},J_{2})$ plane. 
A change of variable $u=r^2$ reduces the integral over $r$ to the form
(\ref{eq:singularity-computation-function1})
and we conclude that the leading singularity is
\begin{equation}
    \phi^{\rm sing}
    = \left\{
      \begin{array}{ll}
          C_{1}(x-x_{0})^{(a_1+a_2)/2} \ln|x-x_{0}| & \\
            \quad +C_{2} (x-x_{0})^{(a_{1}+a_{2})/2} H(x-x_{0})
      & a_{1},a_{2}~\mbox{: even}\\
      0 & \mbox{otherwise}.
     \end{array}
    \right. 
\end{equation}

\paragraph{Case ii), $\lambda_1\lambda_2<0$}: 

Without loss of generality, we may assume,
shifting and scaling $J_{1}$ and $J_{2}$,
that $\lambda_{1}=1~,\lambda_{2}=-1$,
and the saddle is at $(J_{1},J_{2})=(0,0)$.
We need then to study
\begin{equation}
    \varphi(z)
    = \int_{-c_1}^{c_1}\int_{-c_2}^{c_2} \frac{J_{1}^{a_{1}}J_{2}^{a_{2}}}{J_{1}^{2}-J_{2}^{2}-(z-x_{0})}dJ_{1}dJ_{2}.
\end{equation}
The function $\varphi(z)$ is zero except for even $a_{1}$ and $a_{2}$.
We perform the change of variables
\begin{equation}
j_1=J_1+J_2~,~j_2=J_1-J_2~;~dj_1\,dj_2=2\,dJ_1\,dJ_2.
\end{equation}
Forgetting a constant factor and modifying the integration domain (the important thing is to integrate over a neighborhood of $(0,0)$), 
we are led to study
\begin{eqnarray}
    \varphi(z)
   & = & \int_{-c_1}^{c_1}\int_{-c_2}^{c_2} \frac{(j_{1}+j_2)^{a_{1}}(j_1-j_{2})^{a_{2}}}{j_{1}j_{2}-(z-x_{0})}dj_{1}dj_{2} \nonumber \\
   &=& \sum_{k=0}^{a_1}\sum_{l=0}^{a_2} C_{kl} \int_{-c_1}^{c_1}\int_{-c_2}^{c_2} \frac{j_1^{k+l}j_2^{a_1+a_2-k-l}}{j_{1}j_{2}-(z-x_{0})}dj_{1}dj_{2}.
\end{eqnarray}
After a further change of variables
\begin{equation}
u=j_1j_2~,~v=j_2~;~dj_1\,dj_2=\frac{1}{|v|}du\,dv
\end{equation}
we obtain
\begin{eqnarray}
    \varphi(z)
    && = \sum_{k=0}^{a_1}\sum_{l=0}^{a_2} 2C_{kl} \nonumber \\
    && \times \left( \int_{0}^{c_1c_2}du\int_{u/c_1}^{c_2}dv + \int_{-c_1c_2}^{0}du\int_{-u/c_1}^{c_2}dv\right)\frac{u^{k+l}v^{a_1+a_2-2(k+l)}}{(u-(z-x_0))|v|}.
\end{eqnarray}
Remembering that $a_1+a_2$ is even, we perform the integration over $v$:
\begin{eqnarray}
    && \varphi(z) =  \sum_{k=0}^{a_1}\sum_{l=0}^{a_2} 2C_{kl} \int_{-c_1c_2}^{c_1c_2}\frac{u^{k+l}}{(u-(z-x_0))}du \nonumber \\
    && \times \left\{
      \begin{array}{lc}
          \frac{1}{\gamma}(c_2^{\gamma}-(u/c_1)^{\gamma}), & \gamma=a_1+a_2-2(k+l)\neq 0\\
          \ln|c_2|- \ln|u/c_1|, & \gamma=a_1+a_2-2(k+l)=0.
      \end{array} 
    \right.
\end{eqnarray}
The terms with $2(k+l)\neq a_1+a_2$ are not singular, so we are left with 
\begin{equation}
\sum_{k+l=(a_1+a_2)/2} C_{kl}\int_{-c}^{c}\frac{u^{k+l}\ln|u|}{u-z}du.
\end{equation}
According to \ref{sec:main_computation} again,
the singularity of $\phi(x)$ is then
\begin{equation}
    \phi^{\rm sing}(x)
    = C_{1} (x-x_{0})^{(a_{1}+a_{2})/2}\ln|x-x_{0}|
    + C_{2} (x-x_{0})^{(a_{1}+a_{2})/2} H(x-x_{0})
\end{equation}
for $a_{1}$ and $a_{2}$ even.

\section{Singularity computation}
\label{sec:main_computation}

We first show that the function
\begin{equation}
    \label{eq:nosingularity-computation}
    \phi(x) = \lim_{y\to 0^{+}} \int_{-c}^{c} \frac{u^{\alpha}}{u-(x+iy)}du,
    \quad \alpha\in\mathbb{Z}_{+}
\end{equation}
does not have any singularity.
Then we compute singularities of the functions
\begin{equation}
    \label{eq:singularity-computation-function1}
    \phi(x) = \lim_{y\to 0^{+}} \int_{0}^{c} \frac{u^{\alpha}}{u-(x+iy)}du,
    \quad \alpha\in \mathbb{R}\setminus \{\cdots,-3,-2,-1\},
\end{equation}
and
\begin{equation}
    \label{eq:singularity-computation-function2}
    \phi(x) = \lim_{y\to 0^{+}} \int_{-c}^{c} \frac{u^{\alpha}\ln|u|}{u-(x+iy)} du,
    \quad \alpha\in\mathbb{Z}_{+}.
\end{equation}
The constant $c$ is assumed to be positive in the above three functions.
All the singularities found in this paper reduce to a computation
of the singularity close to $x=0$ of one of the above two functions,
(\ref{eq:singularity-computation-function1})
and (\ref{eq:singularity-computation-function2}).
We show in this appendix that singular parts of $\phi$ are
\begin{equation}
    \label{eq:singularity-function1a}
    \phi^{\rm sing}(x)
    = - x^{\alpha} \ln|x| + i\pi x^{\alpha} H(x),
    \quad \alpha\in\mathbb{Z}_{+}
\end{equation}
and
\begin{equation}
    \label{eq:singularity-function1b}
    \phi^{\rm sing}(x)
    = C_{1}x^{\alpha}H(x) + C_{2}(-x)^{\alpha}H(-x)
    + i\pi x^{\alpha} H(x),
    \quad \alpha\in\mathbb{R}\setminus\mathbb{Z}
\end{equation}
for the function (\ref{eq:singularity-computation-function1}),
and
\begin{equation}
    \label{eq:singularity-function2}
    \phi^{\rm sing}(x)
    = Cx^{\alpha}H(x) + i\pi x^{\alpha}\ln|x|,
    \quad \alpha\in\mathbb{Z}_{+}
\end{equation}
for the function (\ref{eq:singularity-computation-function2}).

\subsection{No singularity of the function (\ref{eq:nosingularity-computation})}
We start from the function (\ref{eq:nosingularity-computation})
\begin{eqnarray}
    \phi(x)
    &=& \lim_{y\to 0+} \left[ 
      \int_{-c}^{c} \frac{(u-x)u^{\alpha}}{(u-x)^2+y^2}du
      +iy\int_{-c}^{c}\frac{u^{\alpha}}{(u-x)^2+y^2}du \right].
\end{eqnarray}
We denote the real and the imaginary parts of $\phi(x)$ by
$\phi_{R}(x)$ and $\phi_{I}(x)$ respectively.
The imaginary part $\phi_{I}(x)$ can be computed using the change
of variable $s=(u-x)/y$:
\begin{equation}
    \phi_{I}(x) = \pi x^{\alpha} ( H(x+c) - H(x-c) ),
\end{equation}
where $H$ is the Heaviside step function.
Thus $\phi_{I}(x)$ has no singularity around $x=0$.

The real part is simply:
\begin{eqnarray}
    \phi_{R}(x)
    &=& PV \int_{-c}^{c} \frac{u^{\alpha}}{u-x} du \nonumber\\
    &=& PV \int_{-c}^{c} \frac{x^{\alpha}}{u-x} du
    + \sum_{l=0}^{\alpha-1} C_{l,\alpha} x^{l} 
    \int_{-c}^{c} (u-x)^{\alpha-l-1} du
    \label{eq:no-singularity-real}
\end{eqnarray}
where $PV$ denotes the principal value.
The sum on the right-hand-side is clearly regular,
and the first term gives:
\begin{equation}
    \label{eq:no-singularity}
    PV\int_{-c}^{c} \frac{x^{\alpha}}{u-x} du
    = x^{\alpha} (\ln|c-x| - \ln|-c-x|),
\end{equation}
so that no singularity appears around $x=0$.

\subsection{Singularity of the function (\ref{eq:singularity-computation-function1})}

We study the function (\ref{eq:singularity-computation-function1}),
which differs from the function (\ref{eq:nosingularity-computation})
in the lower bound of the integral.
The imaginary part, $\phi_{I}$, is then:
\begin{equation}
    \phi_{I}(x) = \pi x^{\alpha} ( H(x) - H(x-c) ),
\end{equation}
and hence the singularity of $\phi_{I}$ around $x=0$ is
\begin{equation}
    \phi_{I}^{\rm sing}(x) = \pi x^{\alpha}H(x).
\end{equation}
The real part is also directly obtained using
equation (\ref{eq:no-singularity}):
\begin{equation}
    \phi_{R}(x) = x^{\alpha} (\ln|c-x| - \ln|-x|)
\end{equation}
for $\alpha\in\mathbb{Z}_{+}$.
Thus, the singularity of $\phi_{R}$ around $x=0$ is
\begin{equation}
    \phi_{R}^{\rm sing}(x) = - x^{\alpha}\ln|x|,
    \quad \alpha\in\mathbb{Z}_{+}.
\end{equation}
Combining real and imaginary part gives (\ref{eq:singularity-function1a}).

The final step for the function (\ref{eq:singularity-computation-function1})
is to investigate the real part
for $\alpha\in\mathbb{R}\setminus\mathbb{Z}$.
We divide $\phi_{R}(x)$ in two parts:
\begin{equation}
    \label{eq:phiR-principal}
    \phi_{R}(x)
    = \lim_{\epsilon\to 0} \left(
      \int_{0}^{|x|-\epsilon} \frac{u^{\alpha}}{u-x} du
      + \int_{|x|+\epsilon}^{c}\frac{u^{\alpha}}{u-x} du \right).
\end{equation}
This division is just the definition of the principal value for $x>0$,
and is also valid for $x<0$.
We then use the expansions, for $u<|x|$ and $u>|x|$ respectively:
\begin{equation}
    \label{eq:expansion-u-x}
    \frac{1}{u-x} = \frac{-1}{x} \sum_{k=0}^{\infty}
    \left( \frac{u}{x} \right)^{k}
    \quad ; \quad
    \frac{1}{u-x} = \frac{1}{u} \sum_{k=0}^{\infty}
    \left( \frac{x}{u} \right)^{k}.
\end{equation}
Substituting the above expressions into (\ref{eq:phiR-principal})
and remembering $\alpha\in\mathbb{R}\setminus\mathbb{Z}$,
we have
\begin{eqnarray}
    \phi_{R}(x)
    &=& \left\{
      \begin{array}{ll}
          - x^{\alpha} \sum_{k=0}^{\infty}
            \left(
              \frac{1}{\alpha+k+1}
              + \frac{1}{\alpha-k} \right)
            + \sum_{k=0}^{\infty} x^{k} \frac{c^{\alpha-k}}{\alpha-k}
          & x>0, \\
          (-x)^{\alpha} \sum_{k=0}^{\infty}
            \left(
              \frac{(-1)^{k}}{\alpha+k+1}
              + \frac{(-1)^{k+1}}{\alpha-k} \right)
            + \sum_{k=0}^{\infty}  x^{k} \frac{c^{\alpha-k}}{\alpha-k}
          & x<0. \\
      \end{array}
    \right.
\end{eqnarray}
Note that all series converge.
The second series for $x>0$ and $x<0$ exactly coincide, hence they do not
contribute any singularity.
The singularity comes from the first series and is: 
\begin{equation}
    \phi_{R}^{\rm sing}(x)
    = C_{1} x^{\alpha} H(x) + C_{2} (-x)^{\alpha} H(-x),
    \quad \alpha\in\mathbb{R}\setminus\mathbb{Z}.
\end{equation}
Consequently, we have proved (\ref{eq:singularity-function1b})
for the function (\ref{eq:singularity-computation-function1}).

\subsection{Singularity of the function (\ref{eq:singularity-computation-function2})}
The imaginary part of the function (\ref{eq:singularity-computation-function2}),
denoted by $\phi_{I}(x)$, is
\begin{equation}
    \phi_{I}(x) = \pi x^{\alpha} \ln|x| ( H(x+c) - H(x-c)),
\end{equation}
and the singularity around $x=0$ is
\begin{equation}
    \label{eq:phi-function2-imaginary}
    \phi_{I}^{\rm sing}(x) = \pi x^{\alpha} \ln|x|.
\end{equation}
To compute the real part,
we rewrite the integrand as done in (\ref{eq:no-singularity-real}):
\begin{equation}
    \phi_{R}(x)
    = PV \int_{-c}^{c} \left[ \frac{x^{\alpha}}{u-x} 
      + \sum_{l=0}^{\alpha-1} {}_{\alpha}C_{l}
      (u-x)^{\alpha-l-1} x^{k} \right] \ln|u|
    du.
\end{equation}
We remark that the integral
\begin{equation}
    \int_{-c}^{c} u^{n} \ln|u| du \quad (n\in\mathbb{Z}_{+})   
\end{equation}
can be performed in the sense of improper integral,
and converges to a finite value.
Thus, concentrating on the singularity,
the real part is reduced to
\begin{eqnarray}
    \phi_{R}(x)
    &=& x^{\alpha}~ PV \int_{-c}^{c} \frac{\ln|u|}{u-x} du \nonumber\\
    &=& x^{\alpha} ~ \lim_{\epsilon\to 0} \left(
      \int_{-|x|+\epsilon}^{|x|-\epsilon}
      + \int_{-c}^{-|x|-\epsilon}
      + \int_{|x|+\epsilon}^{c} \right) \frac{\ln|u|}{u-x} du.
\end{eqnarray}
Using expansion (\ref{eq:expansion-u-x}), we obtain
\begin{equation}
    \label{eq:phiR-inside}
    \int_{-|x|+\epsilon}^{|x|-\epsilon}\frac{\ln|u|}{u-x} du
    = \sum_{k=0}^{\infty} \frac{-2}{x^{2k+1}}\left[
      \frac{(|x|-\epsilon)^{2k+1}}{2k+1}\ln| |x|-\epsilon|
      - \frac{(|x|-\epsilon)^{2k+1}}{(2k+1)^{2}}
    \right]
\end{equation}
and
\begin{eqnarray}
    \label{eq:phiR-outside}
    &&\left( \int_{-c}^{-|x|-\epsilon} + \int_{|x|+\epsilon}^{c} \right)
    \frac{\ln|u|}{u-x} du \nonumber\\
    &\simeq & \sum_{k=0}^{\infty} 2x^{2k+1} \left(
      \frac{(|x|+\epsilon)^{-(2k+1)}}{2k+1}\ln||x|+\epsilon|
      + \frac{(|x|+\epsilon)^{-(2k+1)}}{(2k+1)^{2}} \right),
\end{eqnarray}
where we have omitted regular functions in (\ref{eq:phiR-outside}).
The logarithmic terms are cancelled by
adding (\ref{eq:phiR-inside}) and (\ref{eq:phiR-outside})
and taking the limit $\epsilon\to 0$; hence the singularity of $\phi_{R}(x)$ is:
\begin{equation}
    \label{eq:phi-function2-real}
    \phi_{R}^{\rm sing}(x)
    = C x^{\alpha} H(x).
\end{equation}
The imaginary part (\ref{eq:phi-function2-imaginary})
and the real part (\ref{eq:phi-function2-real})
prove (\ref{eq:singularity-function2})
for the function (\ref{eq:singularity-computation-function2}).

\subsection{Relative sign between modes $\bm$ and $-\bm$} 
Coming back to the functions $F$'s and $G$'s,
we are interested in singularities of the function
\begin{equation}
    \label{eq:varphi_m}
    \varphi(z;\bm)
    = \int \frac{\nu(\bJ)}{\bm\cdot\bOmega(\bJ)-z} d\bJ.
\end{equation}
To discuss a possible cancellation between the modes $\bm$ and $-\bm$,
we show a relation between the above function and
\begin{equation}
    \varphi(z;-\bm)
    = \int \frac{\nu(\bJ)}{-\bm\cdot\bOmega(\bJ)-z} d\bJ.
\end{equation}
We note that the numerator $\nu(\bJ)$ may also depend on $\bm$;
in this case one would have to discuss the sign of the numerator separately;
we ignore this dependence in this section.

Remembering that the singularity of the function (\ref{eq:varphi_m})
results in one of the two types of functions
(\ref{eq:singularity-computation-function1})
and (\ref{eq:singularity-computation-function2}),
we consider the relation between 
\begin{equation}
    \phi^{+}_{x_{0}}(x) = \lim_{y\to 0+}\int \frac{h(u)}{u-(x-x_{0})-iy} du
\end{equation}
and
\begin{equation}
    \phi^{-}_{-x_{0}}(x)     = - \lim_{y\to 0+} \int \frac{h(u)}{u+(x+x_{0})+iy} du
\end{equation}
where $h(u)$ is
$u^{\alpha}~(\alpha\in\mathbb{Z}_{+}\cup(\mathbb{R}\setminus\mathbb{Z})$
or $u^{\alpha}\ln|u|~(\alpha\in\mathbb{Z}_{+})$,
and $x_{0}=\bm\cdot\bOmega(\bJ^{\ast})\in\mathbb{R}$ 
with a special point $\bJ=\bJ^{\ast}$.
$\phi^{-}_{-x_{0}}$ is obtained by changing the signs
of the prefactor, $x$ and $y$ in $\phi^{+}_{x_{0}}$,
thus the singular points of $\phi^{+}_{x_{0}}(x)$ and $\phi^{-}_{-x_{0}}(x)$
are $x_{0}$ and $-x_{0}$ respectively.
If $x_{0}\neq 0$, no cancellation occurs in general
due to the different singular points.
If $x_{0}=0$, a cancellation may occur depending
on the relative sign of the singularities
for the modes $\bm$ and $-\bm$.
The change of sign of $y$ implies
the change of sign of the imaginary part, and we use $H(-x)=1-H(x)$.
Hence the relation between the singularities of $\phi^{+}_{0}(x)$
and $\phi^{-}_{0}(x)$ is:
\begin{equation}
    \label{eq:singularity-relation-1}
    \phi_{0}^{-,\rm sing}(x)
    = (-1)^{1+\alpha}\phi_{0}^{+,\rm sing}(x), 
    \quad \alpha\in\mathbb{Z}_{+}
\end{equation}
for the singularity (\ref{eq:singularity-function1a}), and
\begin{equation}
    \label{eq:singularity-relation-2}
    \phi_{0}^{-,\rm sing}(x)
    = (-1)^{\alpha}\phi_{0}^{+,\rm sing}(x),
    \quad \alpha\in\mathbb{Z}_{+}
\end{equation}
for the singularity (\ref{eq:singularity-function2}).
The relations (\ref{eq:singularity-relation-1})
and (\ref{eq:singularity-relation-2}) imply that,
for instance, singularities of $\varphi(z;\bm)+\varphi(z;-\bm)$
and of $\varphi(z;\bm)-\varphi(z;-\bm)$ cancel respectively
if $\alpha$ is $0$ or positive even and $x_{0}=0$.
There is no simple relation such as (\ref{eq:singularity-relation-1}) for 
singularity (\ref{eq:singularity-function1b}), thus
no cancellation is expected in general.


\vspace*{2em}
\begin{thebibliography}{99}
    \bibitem{Landau46} Landau L 1946,
  {\it J.~Phys.~USSR}~{\bf 10} 25

   \bibitem{Maslov85} Maslov V P and Fedoryuk M V 1985 {\it Mat. Sb. (N.S.)} {\bf 127(169)} 445 

    \bibitem{Degond86} Degond P 1986
  {\it Trans.~Am.~Math.~Soc.}~{\bf 294} 435 

    \bibitem{Weitzner67} Weitzner H 1967
  {\it Magneto-Fluid and Plasma Dynamics} edited by Grad H
  (American Mathematical Society, Providence R.I.),
  and references therein

    \bibitem{Glassey95} Glassey R and Schaeffer J 1995
  {\it Commun.~Partial Differ.~Equ.}~{\bf 20} 647 

    \bibitem{MouhotVillani} Mouhot C and Villani C 2011 {\it Acta Mathematica}
{\bf 207} 29

    \bibitem{MouhotVillani2} Mouhot C and Villani C 2010
  {\it J.~Math.~Phys.}~{\bf 51} 015204 

    \bibitem{Lin10} Lin S and Zeng C 2011 {\it Comm. Math. Phys.} {\bf 306} 291 

    \bibitem{Kalnajs77} Kalnajs A J 1977
  {\it Astrophysical J.}~ {\bf 212} 637

    \bibitem{Polyachenko81} Polyachenko V L and Shukhman I G 1981
  {\it Soviet Astronomy (Tr.~Astr.~Zhurn.)}~{\bf 25} 533 

    \bibitem{BinneyTremaine} Binney J and Tremaine S
    2008
    {\it Galactic Dynamics Second Edition} (Princeton University Press)

    \bibitem{Jain07} Jain K, Bouchet F and Mukamel D 2007 {\it J. Stat. Mech.} P11008 

    \bibitem{Campa10} Campa A and Chavanis P H 2010 {\it J. Stat. Mech.} P06001 

    \bibitem{Bachelard10} Bachelard R et al. 2011 {\it J. Stat. Mech.} P03022

    \bibitem{Mathur90} Mathur S 1990
  {\it Mon.~Not.~R.~Astron.~Soc.}~{\bf 243} 529
  
    \bibitem{Weinberg94} Weinberg M D 1994
  {\it Astrophysical J.}~{\bf 421} 481
  
    \bibitem{Weinberg00} Vesperini E and Weinberg M D 2000
  {\it Astrophysical J.}~{\bf 534} 598 
  
    \bibitem{BOY10} Barr\'e J, Olivetti A and Yamaguchi Y Y 2010
  {\it J.~Stat.~Mech.}~P08002
  
    \bibitem{Strogatz} Strogatz S H, Mirollo R E and Matthews P C 1992
  {\it Phys.~Rev.~Lett.}~{\bf 68} 2730 

    \bibitem{Smereka} Smereka P 1998
  {\it Physica D}~{\bf 124},104

    \bibitem{Case1960} Case K M 1960
{\it Phys. Fluids}~{\bf 3} 143 
  
    \bibitem{BOY11}  Barr\'e J, Olivetti A and Yamaguchi Y 2011  {\it J. Phys. A} {\bf 44} 405502

    \bibitem{CluttonBrock72} Clutton-Brock M 1972 {\it Astrophysics and Space Science} 
  {\bf 16} 101
  
    \bibitem{Lighthill} Lighthill M J 1958
  {\it Introduction to Fourier Analysis and Generalized Functions}
  (Cambridge University Press)
    
\end{thebibliography}
\end{document}